\documentclass[twocolumn, balance]{aastex631}

\usepackage{graphicx}
\usepackage{amssymb}
\usepackage{natbib}
\usepackage{xspace}
\usepackage{appendix}

\newcommand{\Msun}{M$_\odot$\xspace}

\begin{document}
\graphicspath{{./}{figures/}}

\title{Radio Variability in Recently-Quenched Galaxies: The Impact of TDE or AGN Driven Outflows}

\author[0000-0002-4235-7337]{K. Decker French}
\affiliation{Department of Astronomy, University of Illinois, 1002 W. Green St., Urbana, IL 61801, USA} 

\author[0000-0003-1991-370X]{Kristina Nyland}
\affiliation{U.S. Naval Research Laboratory, 4555 Overlook Avenue Southwest, Washington, DC 20375, USA}

\author[0000-0002-9471-8499]{Pallavi Patil}
\affiliation{William H. Miller III Department of Physics and Astronomy, Johns Hopkins University, Baltimore, MD 21218, USA}


\author[0000-0002-8989-0542]{Kishalay De}
\affiliation{Department of Astronomy and Columbia Astrophysics Laboratory, Columbia University, 550 W 120th St. MC 5246, New York, NY 10027, USA}
\affiliation{Center for Computational Astrophysics, Flatiron Institute, 162 5th Ave., New York, NY 10010, USA}

\author[0000-0001-9584-2531]{Dillon Dong}
\affiliation{National Radio Astronomy Observatory, P.O. Box O, Socorro, NM 87801, USA}

\author[0000-0003-1714-7415]{Nicholas Earl}
\affiliation{Department of Astronomy, University of Illinois, 1002 W. Green St., Urbana, IL 61801, USA} 

\author[0000-0001-9042-965X]{Samaresh Mondal}
\affiliation{Department of Astronomy, University of Illinois, 1002 W. Green St., Urbana, IL 61801, USA} 

\author[0000-0001-7883-8434]{Kate Rowlands}
\affiliation{William H. Miller III Department of Physics and Astronomy, Johns Hopkins University, Baltimore, MD 21218, USA}
\affiliation{AURA for ESA, Space Telescope Science Institute, 3700 San Martin Drive, Baltimore, MD 21218, USA}

\author[0009-0005-1158-1896]{Margaret Shepherd}
\affiliation{Department of Astronomy, University of Illinois, 1002 W. Green St., Urbana, IL 61801, USA} 

\author[0000-0003-1535-4277]{Margaret E. Verrico}
\affiliation{Department of Astronomy, University of Illinois, 1002 W. Green St., Urbana, IL 61801, USA} 

\begin{abstract}

Outflows and jets launched from the nuclei of galaxies emit radio synchrotron emission that can be used to study the impact of accretion energy on the host galaxy. The decades-long baseline now enabled by large radio surveys allows us to identify cases where new outflows or jets have been launched. Here, we present the results of a targeted VLA program observing four post-starburst galaxies that have brightened significantly in radio emission over the past $\sim20$ years. We obtain quasi-simultaneous observations in five bands (1-18 GHz) for each source. We find peaked spectral energy distributions, indicative of self-absorbed synchrotron emission. While all four sources have risen significantly over the past $\sim20$ years in the 1-2 GHz band, two also show clear recent flares in the 2-4 GHz band. These sources are less luminous than typical peaked spectrum radio AGN. It remains unclear whether these sources are low luminosity analogs of the peaked radio AGN from accreted gas, or driven by tidal disruption events with missed optical flares. Regardless of the source of the accreted material, these newly-launched outflows contain sufficient energy to drive the molecular gas outflows observed in post-starburst galaxies and to drive turbulence suppressing star formation.

\end{abstract}

\section{Introduction}

The long time baselines accessible by large radio surveys have opened new windows on radio sources that vary over decades-long timescales. The FIRST (Faint Images of the Radio Sky at Twenty cm; \citealt{Becker1995}) and VLASS (Very Large Array Sky Survey; \citealt{Lacy2020}) surveys combined cover more than 25 years of time separation and probe fundamentally different transient sources than other multi-wavelength searches. 
Jets and outflows driven via accretion onto supermassive black holes will produce radio synchrotron emission, providing a dust-insensitive method to identify active black holes and characterize their impact on their host galaxies. 

Radio variability provides a unique signature of AGN activity robust against dust obscuration \citep[e.g.,][]{Nyland2020}. With extensive archival data from the FIRST and VLASS surveys, we can compare the observations of radio emission from galaxies several decades apart to search for variability. FIRST was conducted between 1993-2011 and the first epoch of VLASS was conducted between 2017-2019. Because of the difference in frequency between each survey, increases in radio emission (AGN observed to ``turn-on" in the radio) are easier to detect than decreases (AGN observed to ``turn-off"). An object with a detection in FIRST at 1.4 GHz and a non-detection in VLASS at 2-4 GHz could simply have a steep radio spectrum. However, objects with non-detections in FIRST and detections in VLASS are either true variable sources, sources with peaked or steep inverted-slope radio spectra indicative of compact sources \citep{Odea2021}, or both.

\citet{Nyland2020} conducted a systematic search for SDSS- and WISE- selected quasars that were radio quiet during FIRST observations and radio loud during VLASS, obtaining follow-up radio observations for 14 candidate variable sources. These sources had both peaked radio spectra and higher 1.4 GHz emission at the time of observation than measured by FIRST, possibly indicating young, compact jets.

Radio variability thus provides a tracer of the energy in outflows of jets launched into the host galaxy, which can be used to place important constraints on the role of AGN feedback. Studies of AGN feedback that focus on luminous AGN are limited by the fact that AGN feedback appears to be complex and not instantaneous. AGN host galaxies often have high star formation rates and large gas supplies \citep[e.g.,][]{Florez2020, Zhuang2021, Ward2022}, though spatially-resolved observations show evidence that AGN may deplete the central molecular gas in their hosts \citep{Ellison2021}. We use an alternative approach, which is to select galaxies which have recently quenched, and to study the interplay between energy sources and the interstellar medium in these quenched hosts.

Recently-quenched galaxies can be identified as they pass through the post-starburst phase by using optical spectroscopy to constrain the recent star formation histories. Several methods have been used to identify galaxies with recent bursts of star formation that dominate the  optical light of the galaxy, yet with low or declining current star formation rates indicating the galaxy has recently quenched or is currently undergoing quenching \citep{Wild2009,Alatalo2016a,French2018,French2021}. Post-starburst galaxies have colors, morphologies, and kinematics consistent with evolving from star-forming to quiescent after a gas-rich major or minor merger \citep[e.g.,][]{Zabludoff1996,Yang2008,Pracy2013,Sazonova2021}. Due to the angular momentum loss required for the gas fueling star formation to reach the nucleus, the peak in AGN activity may be $\gtrsim100$Myr after the peak in starburst activity, comparable to the timescale over which a galaxy will be identifiable as post-starburst \citep{Davies2007, Schawinski2009, Wild2010, Hopkins2012b, Cales2015, Ellison2025}.

AGN activity in post-starburst galaxies traced by X-ray observations \citep{Lanz2022} and optical emission line ratios \citep{Yan2006,Yang2006} shows weak, LINER (Low Ionization Nuclear Emission-line Region)-like emission. However, some post-starburst galaxies may have deeply obscured central nuclei \citep{Smercina2018,Baron2022}, from which short wavelength AGN signatures may be hidden. Furthermore, the variability of AGN from active to quiescent can occur on timescales ($\sim10^4-10^5$ years) much shorter than the post-starburst phase ($\sim 10^8-10^9$ years) \citep{Lintott2009, Schawinski2015, Keel2017, Shen2021}. Extended emission line regions in post-starburst galaxies have shown signs of flickering AGN activity, with evidence of a duty cycle of $\sim1-2\times10^5$ years where the AGN is only ``on" 5\% of the time \citep{French2023b}. Similar AGN fractions have been seen in X-ray observations of higher redshift post-starburst galaxies, where the duty cycle may explain the lack of excess AGN activity in post-starburst galaxies with high velocity outflows \citep{Almaini2025}. It is important to take into account the obscuration and variability of AGN to estimate the true impact of AGN activity during the phase of galaxy quenching.

Another source of radio variability is tidal disruption events (TDEs), where outflows and jets can produce radio synchrotron emission \citep[e.g.,][]{Alexander2020}. The landscape for TDE observations in the radio has evolved significantly in recent years. The extreme TDE Swift J1644 has had a long-lived radio counterpart indicative of a relativistic jet pointed along our line of sight \citep{Berger2012, Zauderer2013, Eftekhari2018}. Weaker radio activity indicative of non-relativistic outflows was seen for other events, notably ASASSN-14li \citep{Alexander2016}, with the radio emission observed shortly after the optical discovery. ASASSN-14li belongs to the significantly more abundant class of non-relativistic, thermal TDEs, discovered in the optical. A similar event, non-relativistic optical TDE ASASSN-15oi initially showed no radio emission, but follow-up observations 6 months after the optical discovery showed a delayed radio flare \citep{Horesh2021}. These observations prompted additional late-time targeted radio observations. Late-time radio flares were then found in several other optical TDEs \citep{Horesh2021b, Cendes2022}, and larger samples now find late-time radio emission in $\sim40$\% of the sample, at times 500-2000 days after the optical peak \citep{Cendes2024, Somalwar2025a, Goodwin2025}. The radio lightcurves of TDEs thus have significant range, the full diversity of which is certainly underexplored. Some events now show multiple radio flares \citep{Goodwin2024} and theoretical models for outflows from disk instabilities predict radio emission could last for 10-20 years \citep{Piro2024}. Given the increased TDE rate in the post-starburst galaxies we target here \citep{Arcavi2014, French2016, French2020}, radio transients in this post-starburst galaxy sample could be from serendipitous TDEs.

In this work, we present the results from a Karl G. Jansky Very Large Array (VLA) program to follow up four post-starburst galaxies with evidence for variable radio emission. We describe the sample in \S\ref{sec:sample}, observations and data in \S\ref{sec:obs}, and results in \S\ref{sec:results}. We discuss the interpretation of these results in \S\ref{sec:discussion} and summarize our conclusions in \S\ref{sec:conclusions}. When needed, we assume a flat $\Lambda$CDM cosmology with $h=0.7$, $\Omega_m = 0.3$.

\section{Sample Selection}
\label{sec:sample}

\begin{figure*}
\begin{center}
\includegraphics[width=\textwidth]{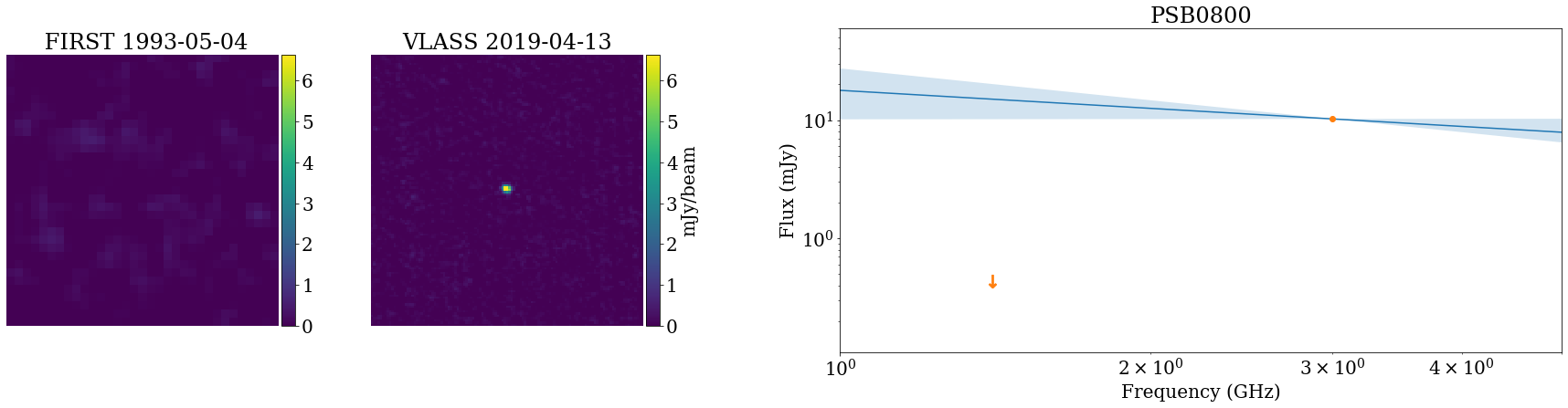}
\includegraphics[width=\textwidth]{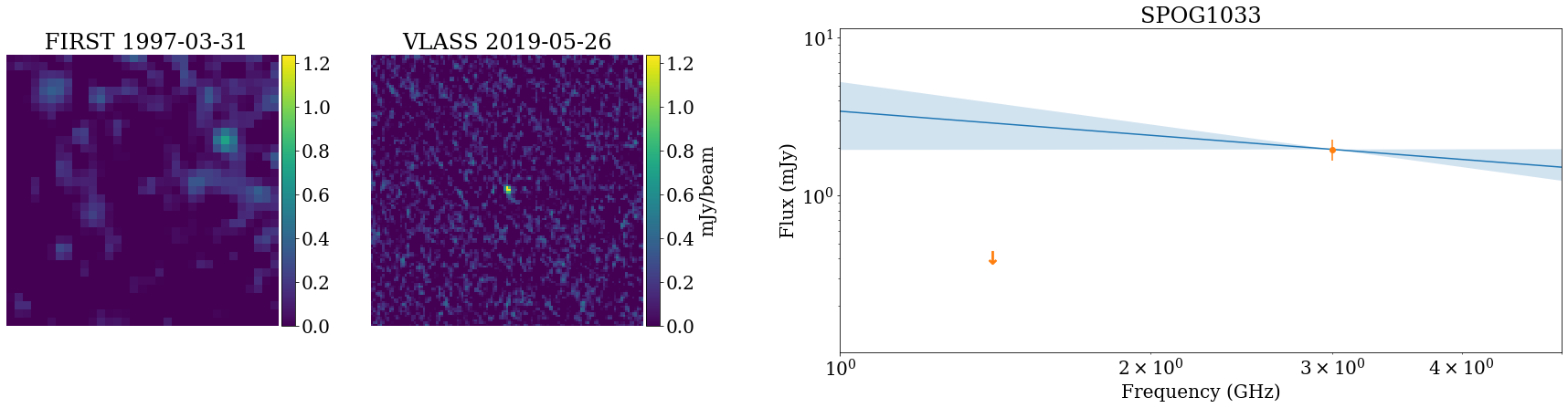}
\includegraphics[width=\textwidth]{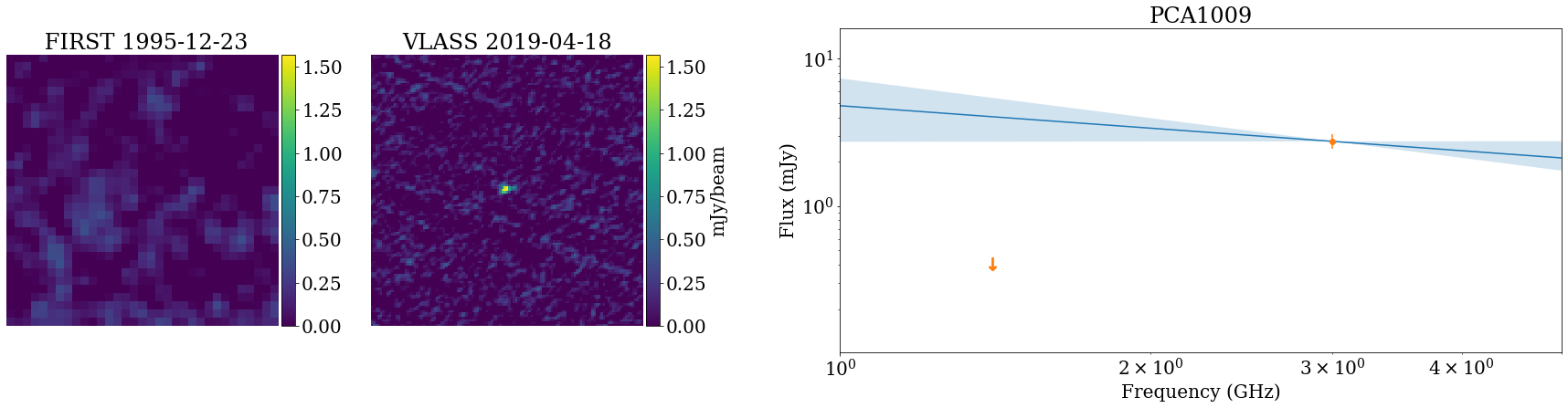}
\includegraphics[width=\textwidth]{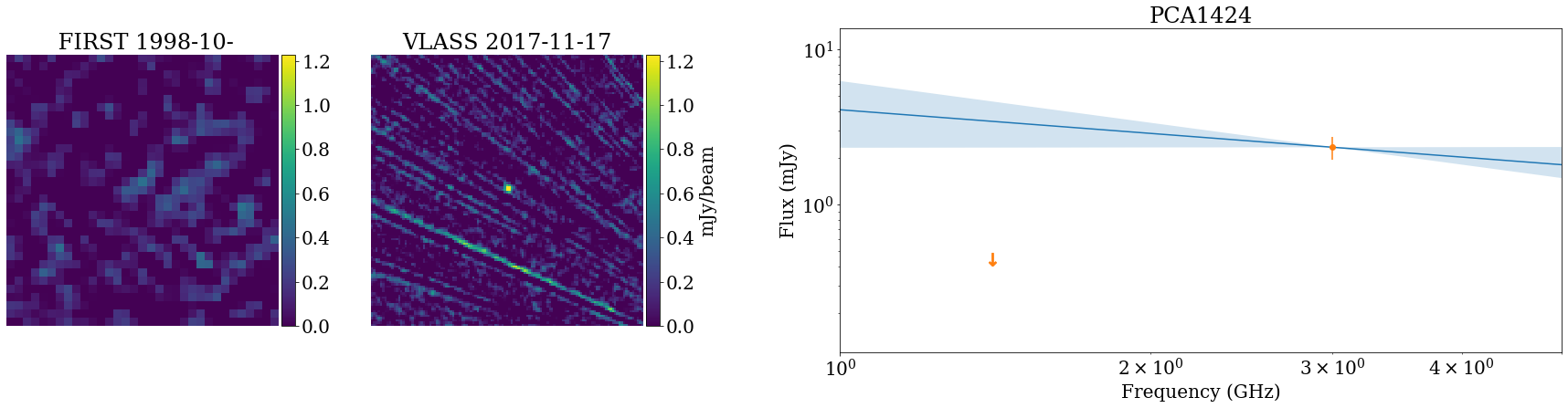}
\end{center}
\caption{Cutout images (1$\times$1 arcmin) from FIRST (left column) and VLASS epoch 1 (middle column) of our four VLA targets, used in sample selection. The right column shows the 3$\sigma$ FIRST upper limits and VLASS epoch 1 detections. The blue shaded region shows the expected spectral energy distribution given the range of power law slopes observed for the sample of post-starburst galaxies with both VLASS and FIRST detections (487 galaxies). The four sources chosen were likely to have risen significantly since the FIRST epochs, despite uncertainties in the spectral slope.
}
\label{fig:selection}
\end{figure*}

We have conducted a similar search as \cite{Nyland2020} for radio-variable AGN in a sample of post-starburst galaxies. We combine three post-starburst samples: PCA selected post-starbursts from \citet{Wild2009}, shocked post-starburst galaxies (SPOGs) from \citet{Alatalo2016a}, and E+A galaxies from \citet{French2018}. All three samples were selected using SDSS single-fiber optical spectroscopy. There is significant overlap between these samples, and they form a combined sample of 5040 galaxies. These selection methods are complementary and capable of selecting quenching galaxies over a wide range of post-burst ages. 

We cross match this post-starburst galaxy sample with the FIRST and VLASS epoch 1 catalogs \citep{Helfand2015, Gordon2020}. While these datasets have different angular resolutions, we expect our sources to be unresolved in both. 14\% of post-starbursts are detected with FIRST and 9\% in VLASS. Non-transient sources may be due to AGN activity, residual star formation or both \citep{Condon1992, Nielsen2012}, but truly variable sources must be driven by nuclear activity. 

231 post-starburst galaxies are detected by FIRST but not VLASS, but given the difference in frequency between these observations, these may not represent truly variable sources. Given the FIRST detection fluxes and VLASS sensitivity limits, the typical power law slope is steeper than $\alpha=-1.8$ if these sources are not variable. These sources could be in the class of Ultra Steep Sources (USSs; \citealt{Bondi2007}), or a combination of steep and variable sources. 

16 post-starburst galaxies are candidate radio-variable AGN, with FIRST non-detections and subsequent detections with VLASS. Some targets may be cases where the radio spectrum is highly inverted due to optically-thick self-absorption, or simply cases where the targets fall just below the detection threshold for FIRST\footnote{For example, this is likely the case for the fourth source shown in Figure \ref{fig:selection}, for which we measure a tentative $2.2\sigma$ detection. }. To exclude these cases, we re-measure the flux for each galaxy in the FIRST observations and require that the inferred radio slope is $\alpha>2.5$, beyond what would be expected for a self-absorbed source \citep{Kellermann1966}. Our final sample has four targets, shown in Table \ref{tab:sample} and Figure \ref{fig:selection}. These four targets are likely variable between the FIRST and VLASS observation epochs, which we proposed to test with new, quasi-simultaneous VLA observations. 

This sample spans a range of the parent post-starburst samples we consider. PSB0800 is selected using an E+A criterion from \citet{French2018}. SPOG1033 is selected from the shocked post-starburst catalog of \citet{Alatalo2016a}. PCA1009 and PCA1424 are selected using the PCA-selected sample of \citet{Wild2009}. While in general there is overlap between these samples (see further discussion of these samples in \citealt{French2023a}), each of these four sources is only present in a single one of the parent samples.

\begin{table*}[]
    \centering
    \begin{tabular}{c c c c c c c}
    \hline
    Galaxy & R.A. (deg) & Decl (deg) & z & $\log M_\star/M_\odot$ & FIRST lim (mJy) & VLASS epoch 1 (mJy) \\ 
    \hline
PSB0800 & 120.06707 & 29.47142 & 0.0453 & 9.98 & $<$0.435 & 10.22 $\pm$ 0.21 \\ 
SPOG1033 & 158.44290 & 47.30110 & 0.0622 & 9.83 & $<$0.408 & 1.97 $\pm$ 0.29 \\ 
PCA1009 & 152.38806 & 23.38215 & 0.0721 & 9.99 & $<$0.408 & 2.75 $\pm$ 0.32 \\ 
PCA1424 & 216.16489 & 21.07683 & 0.0572 & 10.83 & $<$0.447 & 2.34 $\pm$ 0.39 \\ 
\hline
    \end{tabular}
    \caption{Properties of the sample targeted with VLA observations. Coordinates and redshifts (optical convention) are from the SDSS main galaxy sample \citep{sdssdr8}. Stellar masses are from the MPA-JHU catalog \citep{Kauffmann2003, Tremonti2004, Brinchmann2004}. FIRST upper limits are at the 3$\sigma$ level. VLASS epoch 1 fluxes (total) are from the VLASS Quick Look epoch 1 catalog \citep{Gordon2020}. }
    \label{tab:sample}
\end{table*}

\section{VLA Observations}
\label{sec:obs}

\begin{table*}[]
    \centering
    \begin{tabular}{c c c c c}
    \hline
    Galaxy & Date & Band & $\nu$ (GHz) & Total Flux (mJy) \\
    \hline
PSB0800 & 2023 Jul 25 & L & 1.5 & 11.23 $\pm$ 0.06 \\ 
 &  & S & 3.0 & 15.75 $\pm$ 0.04 \\ 
 &  & C & 6.0 & 13.26 $\pm$ 0.02 \\ 
 &  & X & 10.0 & 10.29 $\pm$ 0.02 \\ 
 &  & Ku & 15.0 & 7.62 $\pm$ 0.02 \\ 
\hline
SPOG1033 & 2023 Aug 26 & L & 1.5 & 2.72 $\pm$ 0.04 \\ 
 &  & S & 3.0 & 3.21 $\pm$ 0.03 \\ 
 &  & C & 6.0 & 2.77 $\pm$ 0.03 \\ 
 &  & X & 10.0 & 2.28 $\pm$ 0.02 \\ 
 &  & Ku & 15.0 & 1.85 $\pm$ 0.03 \\ 
\hline
PCA1009 & 2023 Sept 8 & L & 1.5 & 1.03 $\pm$ 0.04 \\ 
 &  & S & 3.0 & 2.96 $\pm$ 0.02 \\ 
 &  & C & 6.0 & 4.37 $\pm$ 0.02 \\ 
 &  & X & 10.0 & 3.75 $\pm$ 0.02 \\ 
 &  & Ku & 15.0 & 2.79 $\pm$ 0.02 \\ 
\hline
PCA1424 & 2023 Jul 21 & L & 1.5 & 1.20 $\pm$ 0.09 \\ 
 &  & S & 3.0 & 2.26 $\pm$ 0.06 \\ 
 &  & C & 6.0 & 2.50 $\pm$ 0.05 \\ 
 &  & X & 10.0 & 1.92 $\pm$ 0.04 \\ 
 &  & Ku & 15.0 & 1.30 $\pm$ 0.06 \\ 
\hline
\hline
    \end{tabular}
    \caption{VLA Observations}
    \label{tab:fluxes}
\end{table*}

We obtained quasi-simultaneous VLA observations of the four targets, through program 23A-292 (PI French), following the observational setup of \citet{Nyland2020}. For each target, we obtained observations in five continuum bands: L: 1–2 GHz, S: 2–4 GHz, C: 4–8 GHz, X: 8–12 GHz, Ku: 12–18 GHz. Each target was observed in successive bands over $\sim1.7$ hours, in order to avoid excessive time separation between observations and obtain a quasi-simultaneous SED. All observations were conducted using the A configuration. 

The data were calibrated and imaged using the VLA calibration and imaging pipelines. We use CASA (Common Astronomy Software Applications, \citealt{casa}) version 2023.1.0.124. Due to the high signal to noise ratio of the data, we also performed self-calibration using the VLA self-calibration pipeline. One galaxy (PCA1424) has a bright nearby radio source, which is not sufficiently masked by the default pipeline. For this galaxy, we manually image and self-calibrate this source using CASA {\tt tclean}. We extract flux measurements for each source using a 2D elliptical Gaussian fit to the image using the CASA task {\tt imfit}. Flux measurements and uncertainties are shown in Table \ref{tab:fluxes}.

For the three epochs of VLASS data, we obtain the single-epoch cutout images from CIRADA\footnote{\url{http://cutouts.cirada.ca/}}. Using the same procedure as for the targeted VLA observations, we use the CASA tool {\tt imfit} to extract flux measurements. For the galaxy with a nearby radio source (PCA1424), we reimage the data using the procedures by \citet{Carlson2022}\footnote{\url{https://github.com/erikvcarlson/VLASS_Scripts}} to split out the needed data from the available measurement sets and image using {\tt tclean}.

\section{Results}
\label{sec:results}

\subsection{Spectral Energy Distributions}
\label{sec:sed}

\begin{figure*}
\begin{center}
\includegraphics[width=0.49\textwidth]{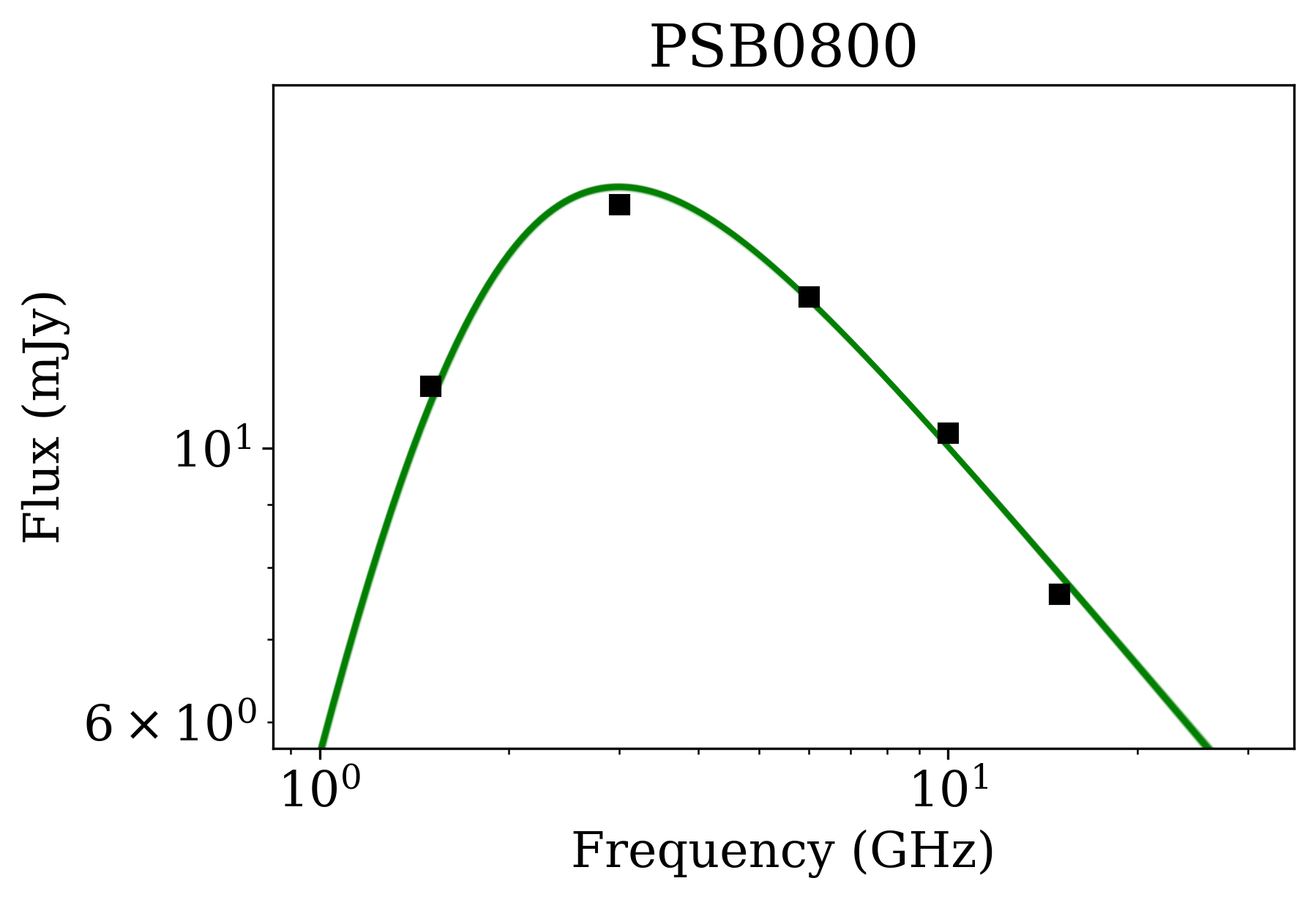}
\includegraphics[width=0.49\textwidth]{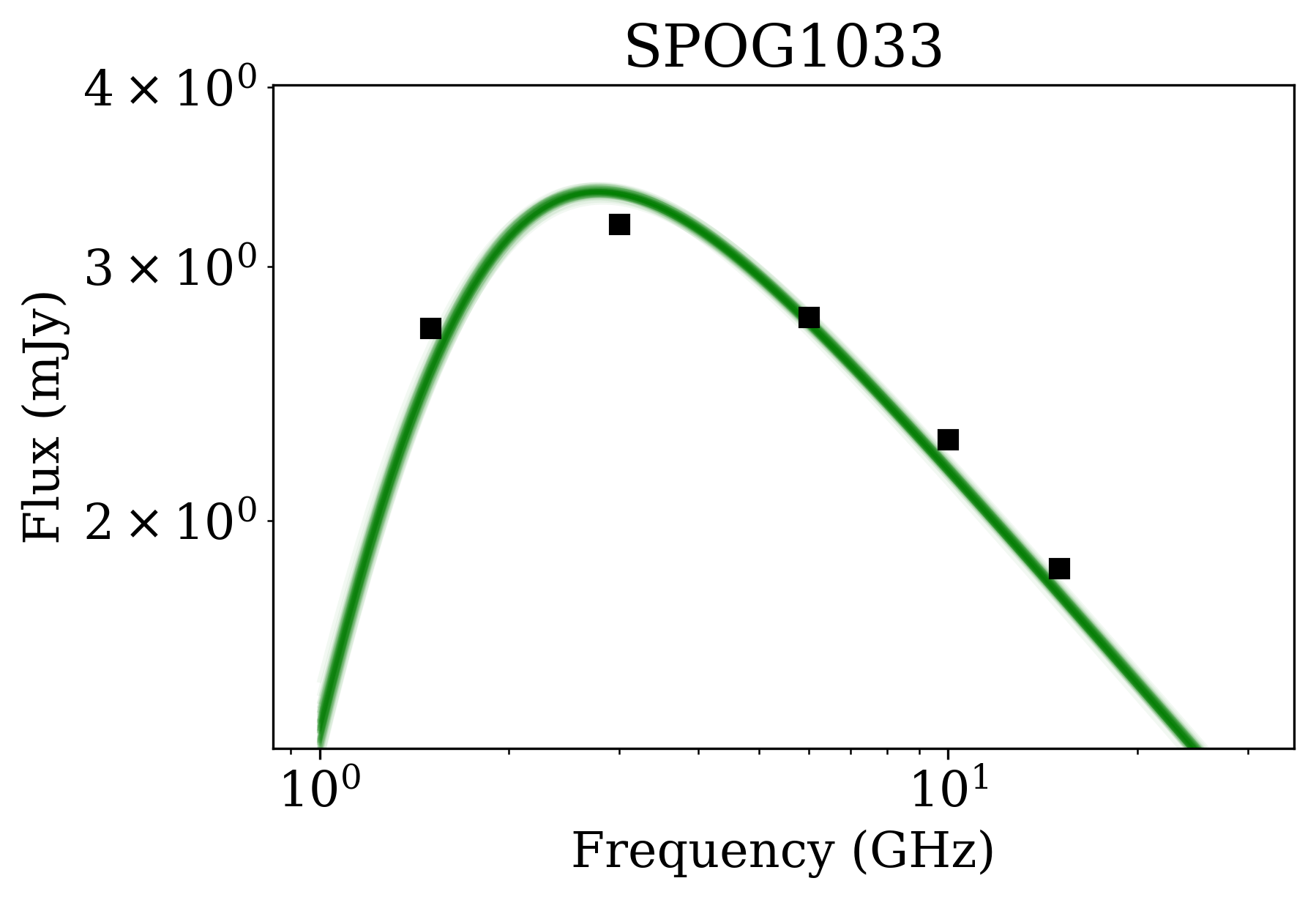}
\includegraphics[width=0.49\textwidth]{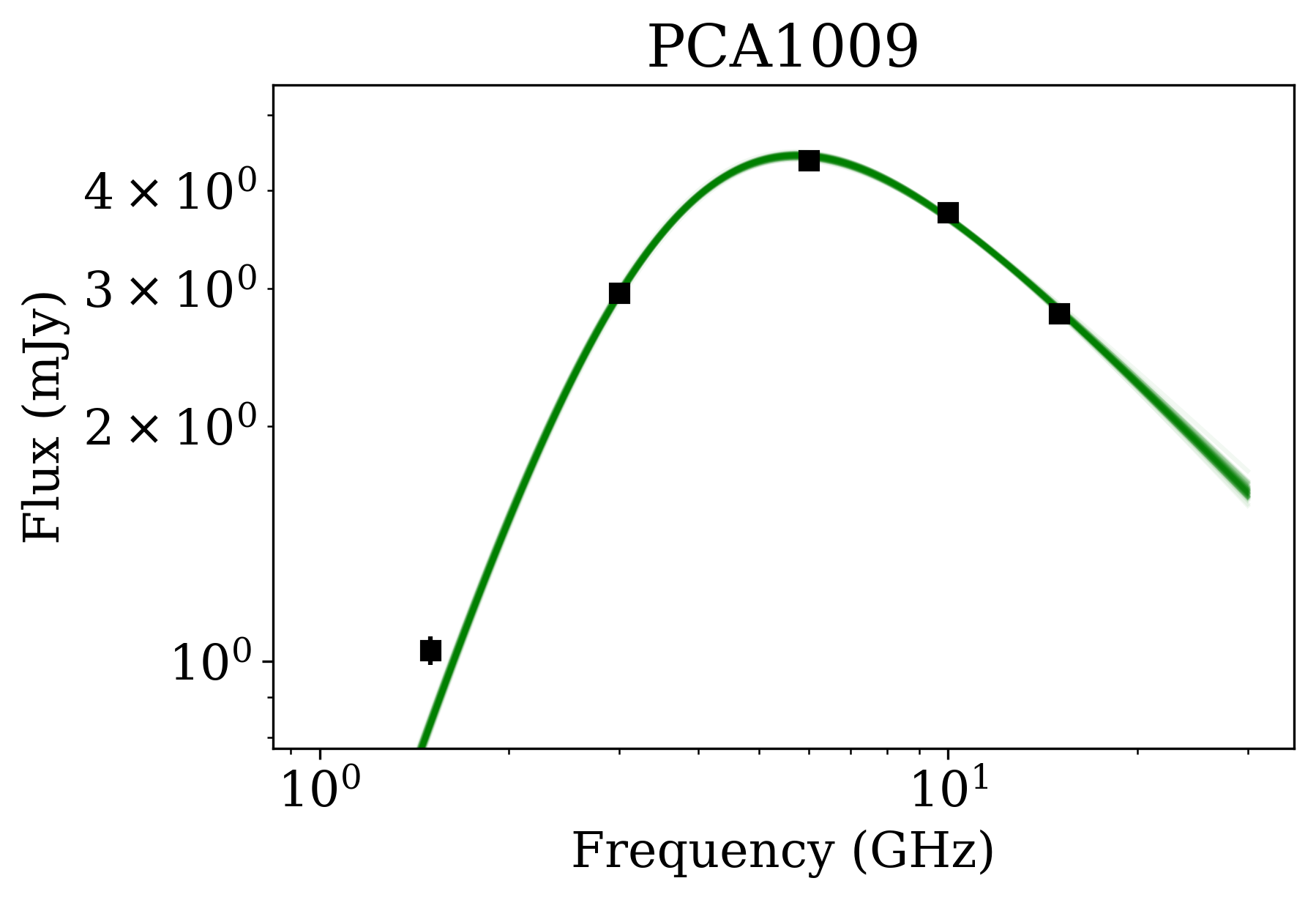}
\includegraphics[width=0.49\textwidth]{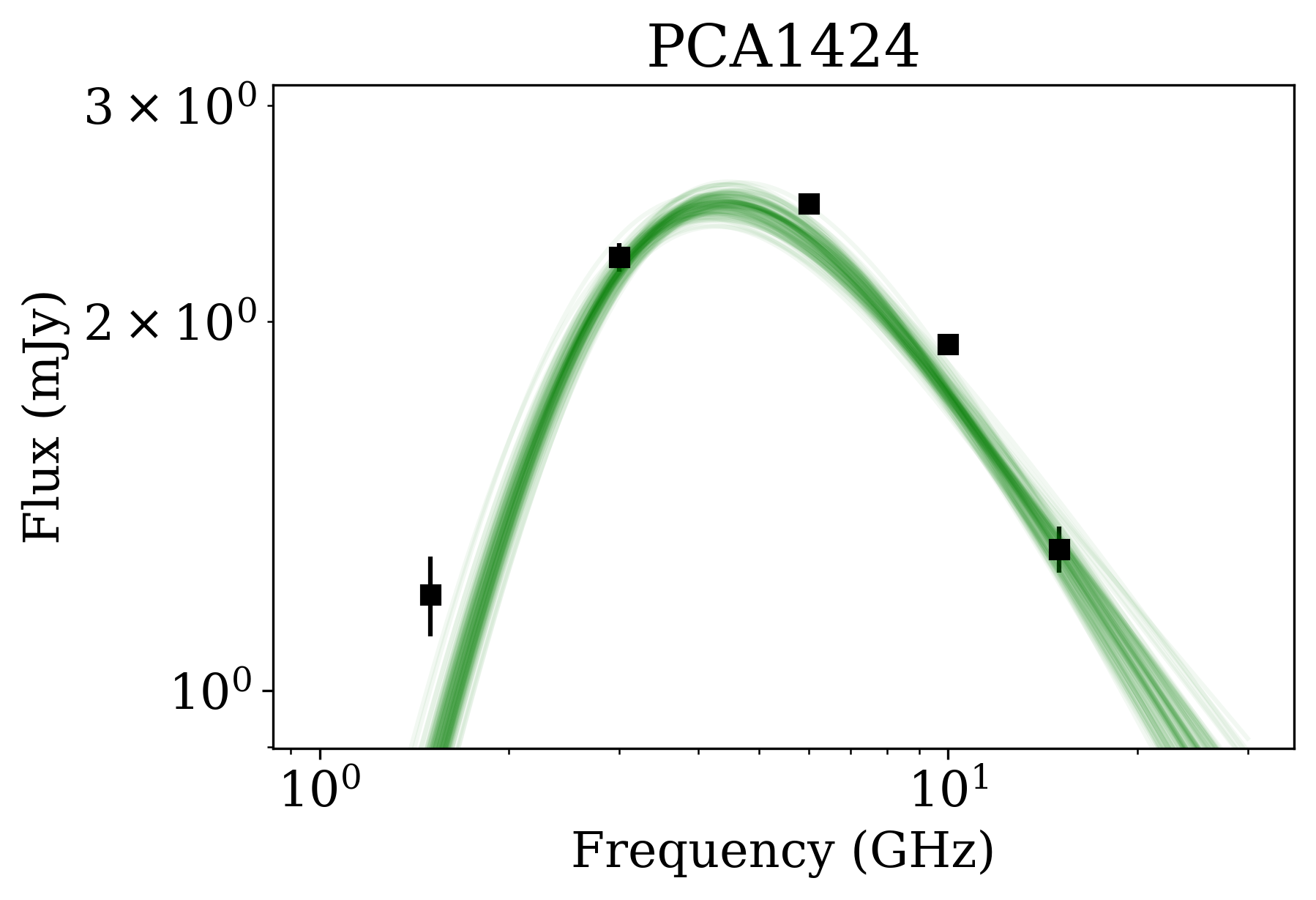}
\end{center}
\caption{Observed SEDs from our targeted VLA observations with best-fit models as discussed in \S\ref{sec:sed}.  
} 
\label{fig:equi}
\end{figure*}

The quasi-simultaneous VLA data allow us to constrain the spectral shape without uncertainties due to the long term variability of these sources. We plot the radio SEDs for the four sources in Figure \ref{fig:equi}. All four sources show peaked SEDs. Two sources have peak frequencies $\sim3$ GHz (PSB0800 and SPOG1033), while the others have higher peak frequencies $\sim6$ GHz (PCA1009 and PCA1424). The observed SED peak and turnover towards lower frequencies indicates that the lower frequency emission is absorbed. In the following analysis, we assume this absorption is due to synchrotron self-absorption (SSA), but we explore this assumption in \S\ref{sec:origin}.

Low frequency data would help to constrain the optically-thick slope. While data taken at a different time than our VLA observations is subject to additional uncertainties due to source variability, we search for archival low frequency data to supplement our SEDs. The LoTSS (LOw-Frequency ARray (LOFAR) Two-metre Sky Survey; \citealt{Shinwell2022}) survey has obtained sensitive 144 MHz observations over a large percentage of the Northern sky. Two of our sources (PSB0800 and SPOG1033) are within the survey footprint, but neither is significantly detected to be included in the LoTSS DR2 source catalog. If we assume a thermal spectral slope of $\alpha=2.5$ from the VLA 1.4 GHz detections, the 144 MHz fluxes should be 38 $\mu$Jy and 9 $\mu$Jy for PSB0800 and SPOG1033, respectively, below the 83 $\mu$Jy/beam median rms sensitivity of LoTSS.

We fit the SEDs using the model of \citet{Granot2002} for the synchrotron emission from a relativistic shock propagating through the interstellar medium, following \citet{Eftekhari2018, Cendes2024}. For a shock propagating through an ambient medium, the electrons will have a power law energy distribution such that $N(\gamma) \propto \gamma^{-p}$ where the Lorentz factor $\gamma$ extends down to a minimum $\gamma_m$, which can vary. This model has been used to fit the peaked spectra of TDEs \citep{Eftekhari2018, Cendes2024} and is analogous to the models used to fit the spectra of peaked radio source AGN \citep{Orienti2012,Patil2022}. 

The slope of each component of the SED will depend on the relative order of the self-absorption frequency $\nu_a$, the synchrotron frequency corresponding to the minimum electron energy in the power law $\nu_m$, and the cooling frequency $\nu_c$. We expect the cooling frequency break from electron cooling to be blueward of our radio SED \citep{Berger2012}, and do not consider the higher frequency portions of the SED here. Thus, we only consider the relative ordering of $\nu_a$ and $\nu_m$ (spectrum 1 and 2 in \citealt{Granot2002}). The well-studied radio-bright TDE Swift J1644 was observed to undergo a change in this relative ordering, due to the differing evolution of $\nu_a$ and $\nu_m$ as a function of time \citep{Berger2012, Zauderer2013, Eftekhari2018}. At early times, $\nu_a < \nu_m$, while at late times $\nu_m < \nu_a$. \citet{Cendes2024} consider the radio SEDs of late-time radio flares from thermal (non-jetted) TDEs, and assume $\nu_m << \nu_a$, for which the resulting SEDs are a good fit to the data. In order to model a smooth transition between the two spectral shapes, \citet{Eftekhari2018} use the ratio of $\nu_a/\nu_m$ to model the spectrum as a weighted average of the two cases. We consider this model for our sources, but in every case find a $\nu_m < \nu_a$ fit is preferred, so we do not alter the model used for the results reported here.

Our model thus consists of three free parameters, a normalization flux $F(\nu_p)$, the self-absorption frequency $\nu_a$, and the power law $p$ which sets the energy distribution of the electrons. Below $\nu_a$, the SED slope is $\beta_1 = 5/2$, and above it is $\beta_2 = (1-p)/2$, with a shape given by
\begin{eqnarray}\begin{array}{rcl}{F}_{\nu } & = & {F}_{\nu }({\nu }_{m})\left[{\left(\displaystyle \frac{\nu }{{\nu }_{m}}\right)}^{2}{e}^{-{s}_{4}{\left(\nu /{\nu }_{m}\right)}^{2/3}}+{\left(\displaystyle \frac{\nu }{{\nu }_{m}}\right)}^{5/2}\right]\\ & & \times {\left[1+{\left(\displaystyle \frac{\nu }{{\nu }_{a}}\right)}^{{s}_{5}({\beta }_{1}-{\beta }_{2})}\right]}^{-1/{s}_{5}},\end{array}\end{eqnarray}
where $s_4 = 3.63p-1.60$ and $s_5 = 1.25-0.18p$. We renormalize each spectrum to a peak flux $F(\nu_p)$. We fit each SED to this model using {\tt emcee} \citep{emcee}. Results from these fits are shown in Figure \ref{fig:equi} and presented in Table \ref{tab:equi}.

If the sources are peaked due to SSA, the source luminosity and size will be related to the peak frequency. If we assume energy equipartition, the minimum magnetic field to produce the observed emission will be $B_{min} \propto (L/R^3)^{2/7}$, where $L$ is the luminosity and $R$ is the source size \citep{Miley1980, Orienti2014, Patil2022}. For self-absorption, the magnetic field is additionally related to the turnover frequency $\nu_p$, as $B_{SSA} \propto \nu_p^5 R^4/L^2$ \citep{Kellerman1981, Orienti2014, Patil2022}. \citet{Orienti2014} demonstrate that $B_{min}\approx B_{SSA}$ for peaked sources. \citet{Patil2022} demonstrates that setting $B_{min}\approx B_{SSA}$ can be used to infer a source size for unresolved sources in peaked AGN sources. 

Following \citet{Cendes2024, Barniol2013}, we incorporate the results of our best-fit peak flux, peak frequency, and $p$ in calculating the energy $E_{eq}$, radius $R_{eq}$ and magnetic field $B_{eq}$ that would produce this spectrum under the assumptions of energy equipartition. We utilize the same assumptions as \citet{Cendes2024} in order to facilitate the comparison with TDEs in \S\ref{sec:tdes}\footnote{We compare our method following \citet{Cendes2024} to the method used by \citet{Patil2022} for AGN to infer source sizes. There are small differences of $\mathcal{O}(1.5\times)$ that result from the differing assumptions of geometry and deviations from equipartition.}. The results of this calculation are shown in Table \ref{tab:equi}.

We note several places where this model is a poor fit to the observed data. The high frequency slope of SPOG1033 would be better fit by a shallower slope, or lower value of $p<2$, which is unphysical in this model. Two sources (PCA1009, PCA1424) with higher peak frequencies show evidence that the low frequency slope is shallower than $\nu^{5/2}$. This behavior is often seen in peaked spectrum AGN \citep{delaParra2024}, from the influence of multiple contributions over a range of source sizes and ages. We discuss the expectations for AGN vs. TDE scenarios further in \S\ref{sec:agn} and \S\ref{sec:tdes}.

\begin{table*}[]
    \centering
    \begin{tabular}{c c c c c c c}
    \hline
    Galaxy & $\nu_{a}$ (GHz) & $F_{\nu_p}$ (mJy) & $p$ & $R_{eq}$ (cm) & $E_{eq}$ (erg) & $B_{eq}$ (G) \\ 
    \hline
PSB0800 & $1.75^{0.01}_{0.01}$ & $16.27^{0.03}_{0.03}$ & $2.20^{0.01}_{0.01}$ & 3.62e+17 & 2.29e+49 & 1.28e-01 \\ 
SPOG1033 & $1.52^{0.02}_{0.02}$ & $3.38^{0.02}_{0.02}$ & $2.00^{0.00}_{0.00}$ & 2.64e+17 & 7.77e+48 & 1.25e-01 \\ 
PCA1009 & $3.73^{0.04}_{0.04}$ & $4.44^{0.02}_{0.02}$ & $2.63^{0.04}_{0.04}$ & 1.43e+17 & 6.93e+48 & 3.16e-01 \\ 
PCA1424 & $2.77^{0.14}_{0.14}$ & $2.48^{0.05}_{0.04}$ & $2.57^{0.15}_{0.14}$ & 1.17e+17 & 2.67e+48 & 2.48e-01 \\ 
\hline
    \end{tabular}
    \caption{Results from Equipartition modeling}
    \label{tab:equi}
\end{table*}

\subsection{Lightcurves}
\label{sec:lightcurves}

\begin{figure*}
\begin{center}
\includegraphics[width=0.49\textwidth]{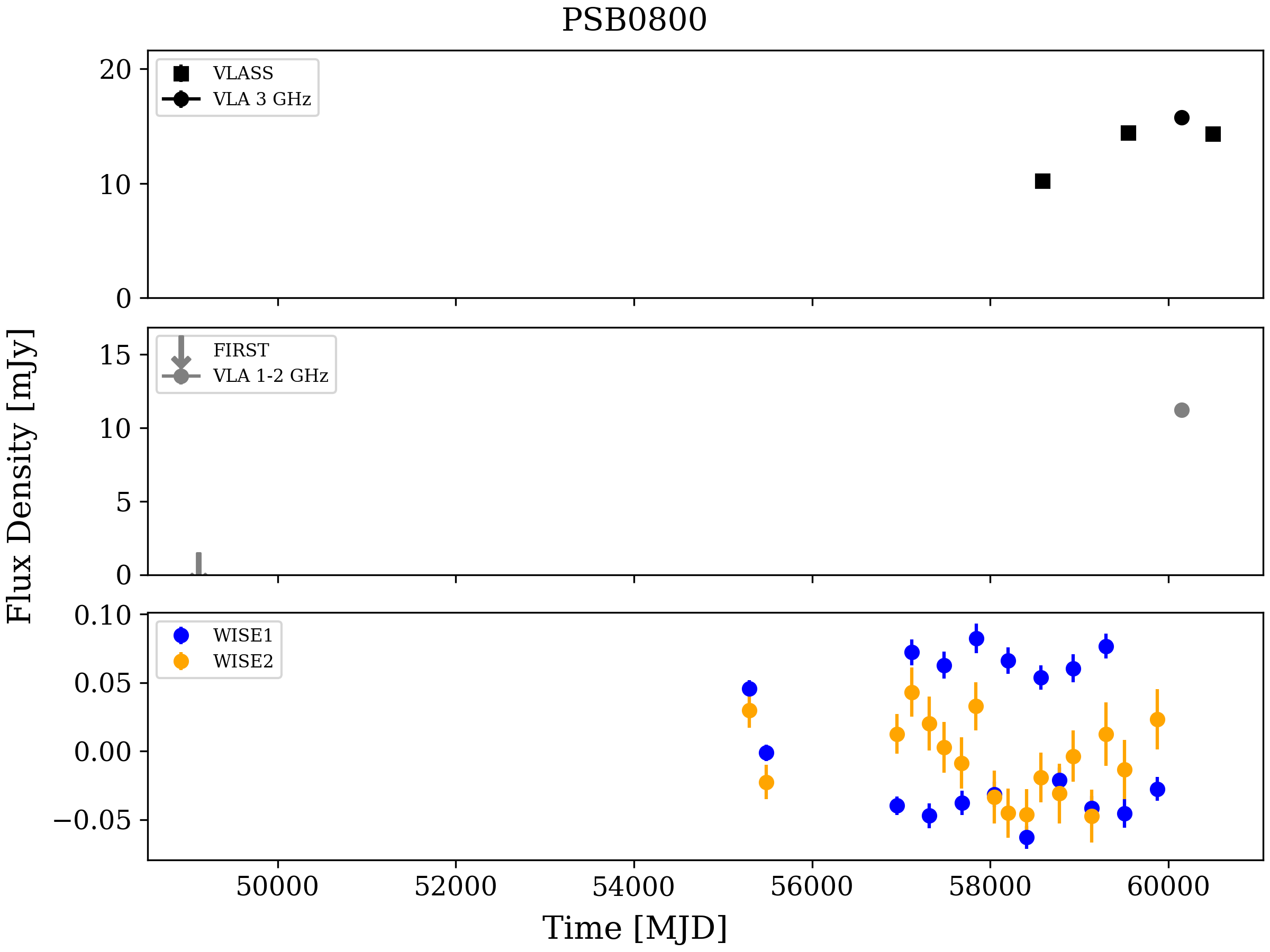}
\includegraphics[width=0.49\textwidth]{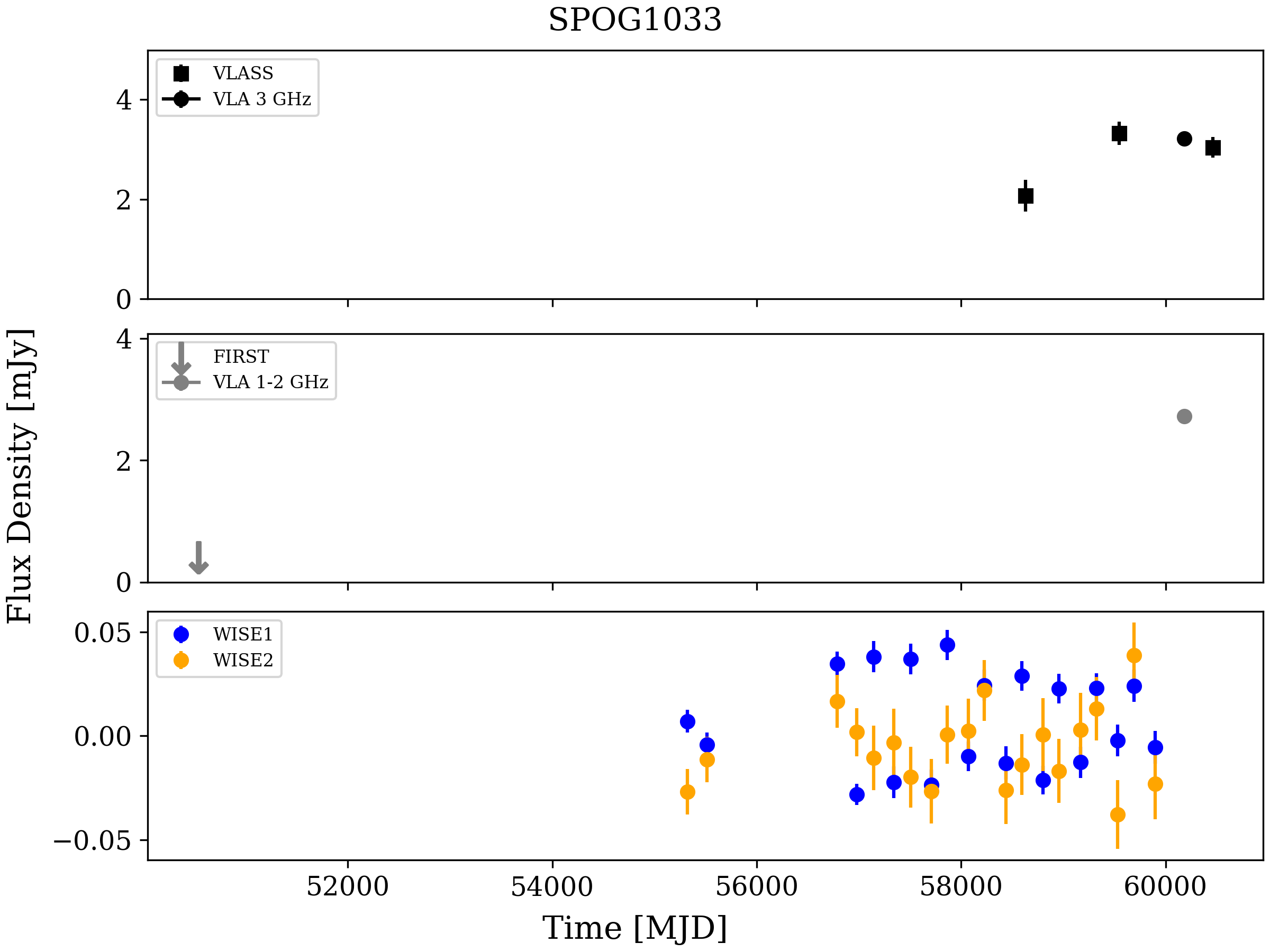}
\includegraphics[width=0.49\textwidth]{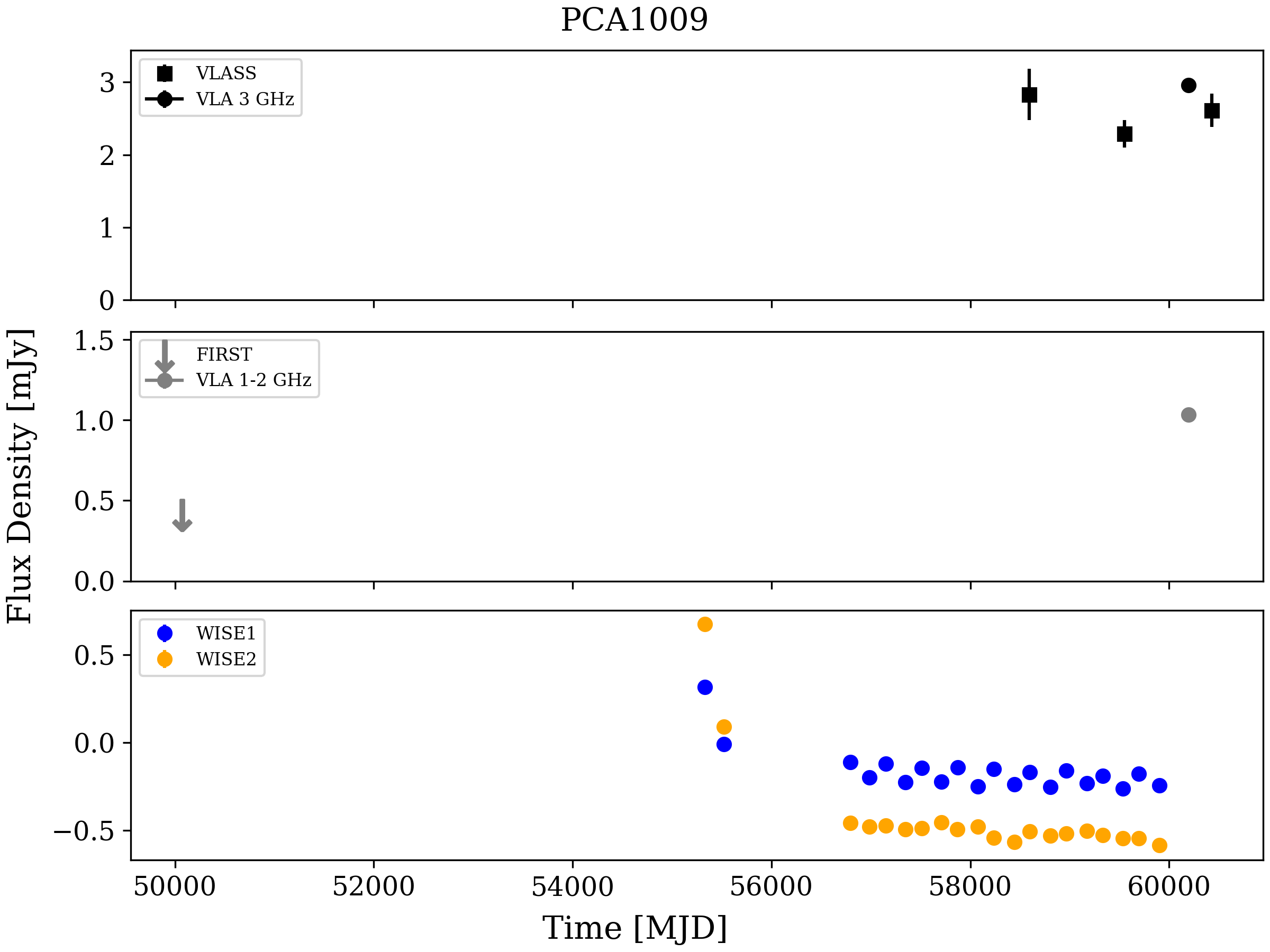}
\includegraphics[width=0.49\textwidth]{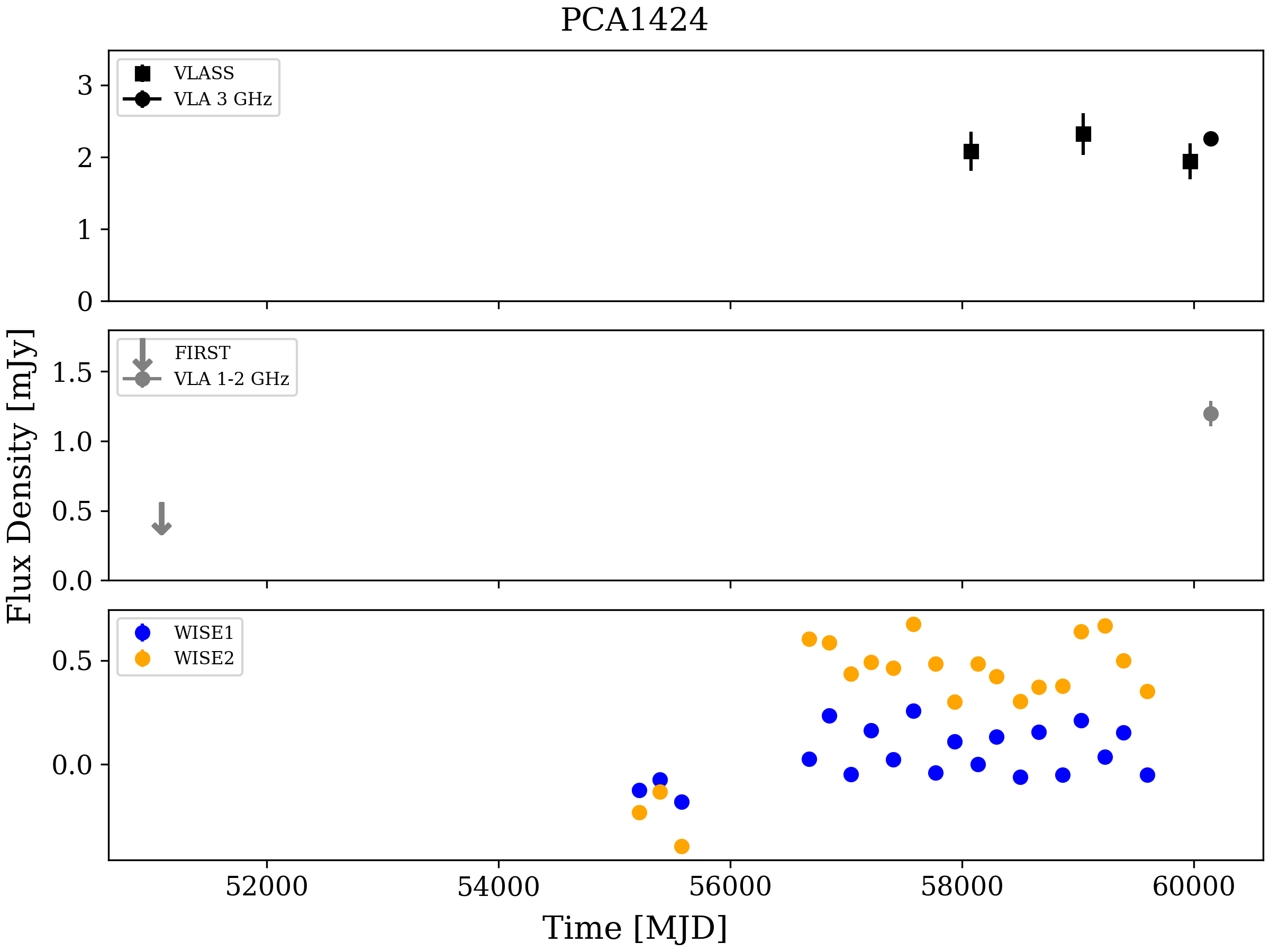}
\end{center}
\caption{Radio and IR lightcurves. (Top) Lightcurves of 3 GHz flux from VLASS (squares) and our targeted VLA observations (circles). The top two sources show clear increases since the first epoch of VLASS, while the bottom two are more constant or variable. (Middle) 1.4 GHz lightcurves from the FIRST upper limits and targeted VLA observations. All four sources have brightened significantly in the L band since the non-detections in FIRST. The four sources have risen by at least $25, 6.6, 2.5, 2.5 \times$ (in order listed in each table, respectively), relative to the FIRST $3\sigma$ upper limits on non-detections. (Bottom) WISE W1 (3.4 $\mu$m) and W2 (4.6 $\mu$m) lightcurves for each source.  No significant flares or trends are seen for two of our targets (PSB0800 and SPOG1033). A clear flare is seen for PCA1009, with a strong decline seen over several years. An IR increase is also seen for PCA1424, though in this case we see an initial increase and plateau, dissimilar to the light echoes seen after transient events. 
}
\label{fig:lc_radio}
\end{figure*}

The targeted VLA observations allow us to test for variability on 17-20 year timescales in the L band (1-2 GHz) and over the past several years in multiple epochs in the S band (2-4 GHz). All four sources have brightened significantly in the L band since the non-detections in FIRST. The four sources have risen by at least $25, 6.6, 2.5, 2.5 \times$ (in order listed in each table, respectively), relative to the FIRST $3\sigma$ upper limits on non-detections. In the S band, we show the lightcurves using both the targeted VLA data and three VLASS epochs in Figure \ref{fig:lc_radio}. Two sources (PSB0800 and SPOG1033) have risen significantly since the first epoch of VLASS and since declined or leveled off. The other two sources (PCA1009, PCA1424) show no coherent trend, with fluxes consistent with the first epoch of VLASS.

We use multi-wavelength archival data to search for variability in optical and IR photometry, for each of the four sources Infrared data from WISE (Wide-field Infrared Survey Explorer; \citealt{Wright2010}) in the W1 (3.4 $\mu$m) and W2 (4.6 $\mu$m) band is available from the all sky WISE survey and the subsequent NEOWISE survey \citep{Mainzer2011}. Using the method from \citet{De2020, Masterson2024}, we measure the differenced photometry in each epoch. IR lightcurves are shown in Figure \ref{fig:lc_radio}. No significant flares or trends are seen for two of our targets (PSB0800 and SPOG1033). A clear flare is seen for PCA1009, with a strong decline seen over several years. This flare has been selected in previous searches for IR flares by \citet{Assef2018} and \citet{Wang2018}, and resembles the dust echoes seen by \citet{Masterson2024}. We discuss the interpretation of this flare further in \S\ref{sec:discussion}. An IR increase is also seen for PCA1424, though in this case we see an initial increase and plateau, dissimilar to the light echoes seen after transient events. The steady increase and WISE color evolution seen for PCA1424 is more similar to the changes seen for some changing-look AGN \citep{Yang2018}.

We use optical data from the following surveys: Pan-STARRS (Panoramic Survey Telescope and Rapid Response System) 3$\pi$ survey \citep{Chambers2016}; ZTF (Zwicky Transient Facility, \citealt{Bellm2019}); ASASSN (All-Sky Automated Survey for Supernovae, \citealt{Shappee2014, Hart2023}); CRTS (Catalina Real-time Transient Survey, \citealt{Drake2009}); and ATLAS (Asteroid Terrestrial-impact Last Alert System, \citealt{Tonry2018, Smith2020-atlas}). All upper limits shown are at the 3$\sigma$ level. For the significant detections, we calculate and subtract the median value to obtain the best estimate for the difference flux in searching for flares. We sigma clip at the $3\sigma$ level over two iterations to remove spurious detections. 

The optical lightcurves are shown in Appendix \ref{sec:appendix} (Figure \ref{fig:optical}). No clear flares or variability are detected in the optical for any of these datasets. However, we note that it is possible an optical transient was missed prior to the first observations we consider here or was dust-obscured (particularly in the cases with IR flares discussed above).

\section{Discussion}
\label{sec:discussion}

\subsection{Origin of the Peaked Radio Spectra}
\label{sec:origin}

The observed SED peak and turnover towards lower frequencies indicates that the lower frequency emission is absorbed. Two possibilities to explain this SED shape are synchrotron self-absorption (SSA) or free-free- absorption (FFA). FFA can generate a range of optically-thick spectral slopes depending on the geometry of the absorbing material \citep{Bicknell1997}. In sources with high densities and ionizing luminosities such as obscured AGN \citep{Patil2022}, FFA has been used to explain the absorption instead of SSA. FFA has also been used to explain the small sizes inferred for low frequency peaked sources \citep{Keim2019}. The optical depth ${\tau }_{\mathrm{ff}}$ due to FFA is given by 
\begin{equation}
{\tau }_{\mathrm{ff}}\approx 3.3\times {10}^{-4}\ {T}_{4}^{-1.35}\ {\nu }_{\mathrm{GHz}}^{-2.1}\ \int {n}_{{\rm{e}}}^{2}\ {{dl}}_{\mathrm{kpc}}
\end{equation}
\citep{Bicknell1997, Patil2022}, where $T_4$ is the temperature (in units of $10^4$ K), ${\nu }_{\mathrm{GHz}}$ is the frequency in GHz, $n_e$ is the electron density in cubic centimeters, and ${l}_{\mathrm{kpc}}$ is the depth in kpc. Using the estimates of the electron density profile from TDE hosts \citep{Alexander2020}, and assuming a temperature of $10^4$ K, the optical depth from FFA at 3 GHz is ${\tau }_{\mathrm{ff}} \sim 10^{-11}-10^{-6}$, too low to explain the turnover frequencies we see in these sources. For a sample of nearby radio AGN with $\lesssim400$ pc resolution integral field spectroscopy studied by \citet{Kakkad2018}, the optical depth from FFA at 3 GHz is at most $\tau=1.5$ for the most extreme case. Our post-starburst sources are likely more similar to the non-AGN sources with TDEs studied by \citet{Alexander2020} than the luminous and deeply obscured sources in a subset of the \citet{Patil2022} sample, and thus SSA is a more likely source of absorption.

Peaked radio AGN are thought to be young sources ($\sim100-1000$ years, \citealt{An2012}). Some of these sources may evolve into large-scale radio AGN, although the rate of compact sources is much greater than the rate of extended sources, meaning some of the compact sources may be short-lived instead \citep[e.g.,][]{An2012,Readhead2024}. If compact sources are short-lived instead of young, they will never evolve into the rarer large-scale sources. The sources we consider here could be young outflows or jets, and we discuss in \S\ref{sec:agn} and \S\ref{sec:tdes} the possibilities that they are driven by AGN or TDEs, respectively. The geometry of these systems is key to interpreting whether these are collimated jets vs. winds, as well as the scale at which energy is being injected into the interstellar medium. The geometry is also important in interpreting the inferred sizes, as we discuss in \S\ref{sec:agn}. Higher spatial resolution observations would be required to directly constrain the geometry of these post-starburst sources.

Another possibility is that these sources could be frustrated jets in dense environments, though we do not see signs of strong attenuation. From the Balmer decrement, 3/4 sources have $A_V \lesssim0.2$ mag, except for PCA1424, which has $A_V\sim1$ mag. Despite the high obscured star formation rates that have been observed in some post-starburst galaxies \cite{Baron2022}, none of the proposed targets have significant IRAS detections, indicating that none of these proposed post-starburst galaxy targets contain obscured starbursts.

\begin{figure*}[t!]
\begin{center}
\includegraphics[width=0.49\textwidth]{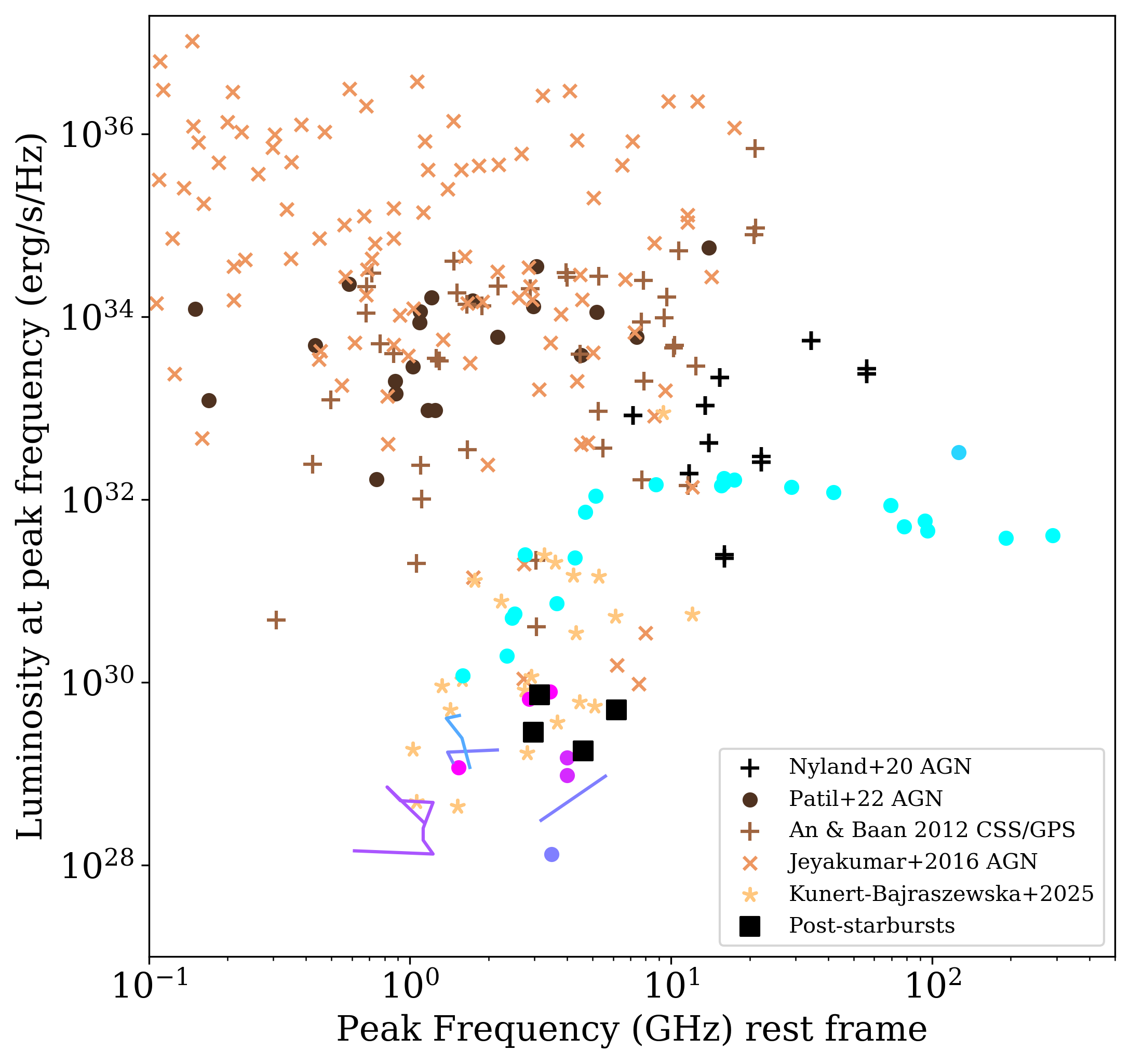}
\includegraphics[width=0.49\textwidth]{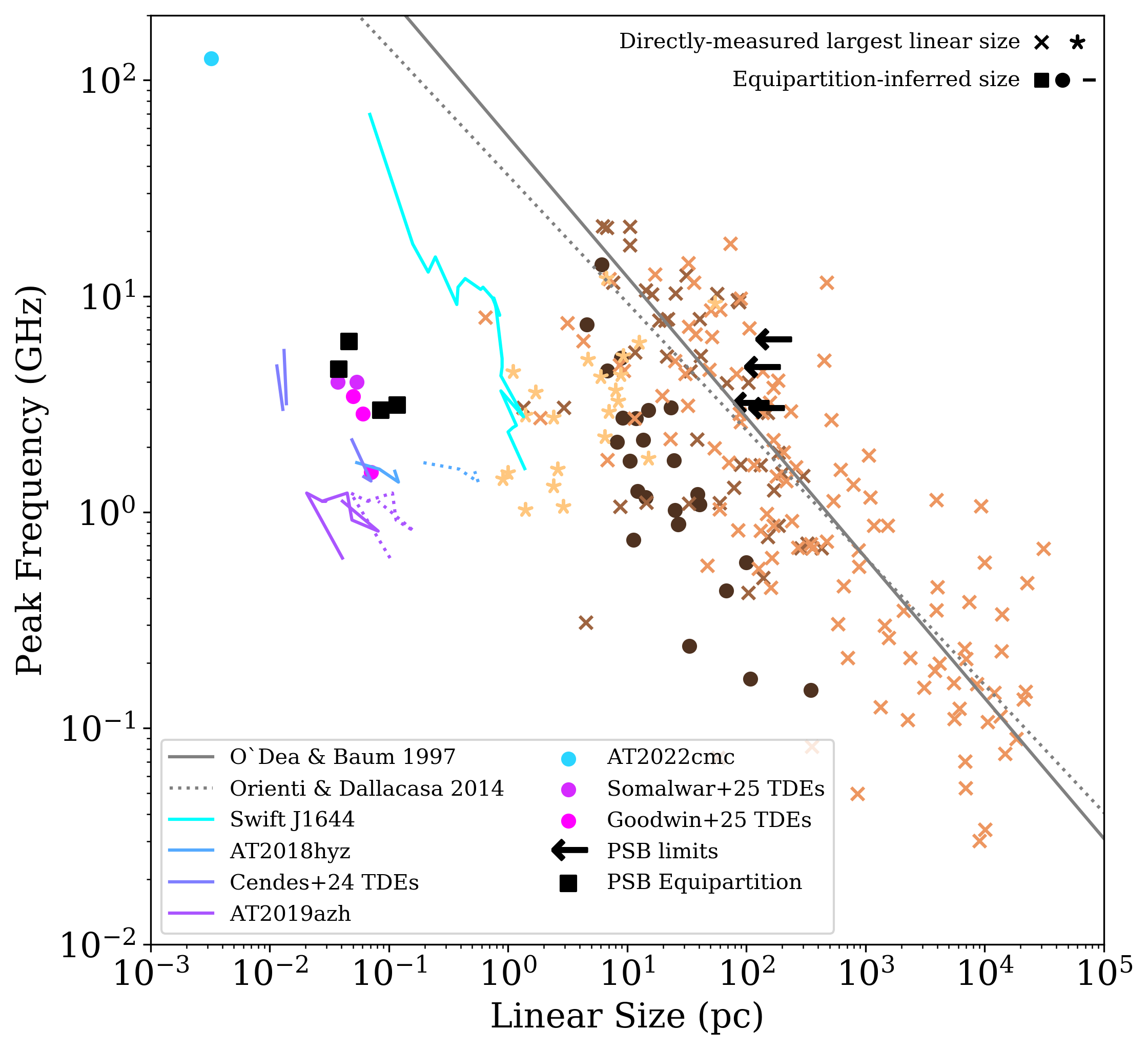}
\end{center}
\caption{Left: Peak frequency vs. luminosity at peak frequency, Right: size vs. rest-frame peak frequency (legend shared). We compare the post-starburst targets to samples of AGN (copper colors; \citet{Patil2022, Kunert2025} and compilations by \citet{An2012}, \citet{Jeyakumar2016}; peak--linear size relations from \citet{Odea1997} and \citet{Orienti2014}) and TDEs (cool colors; relativistic jetted TDEs from Swift J1644 \citep{Eftekhari2018} and AT2022cmc \citep{Pasham2023}, optically-selected events from \citet{Cendes2024, Goodwin2022} (including AT2019azh), the delayed jetted event AT2018hyz \citep{Cendes2022}, radio-identified TDEs from \citealt{Somalwar2025b}), and X-ray identified TDEs from \citealt{Goodwin2025}). For the post-starburst targets, we plot both the limits on the largest size from the VLA Ku-band beam sizes, as well as the inferred sizes from equipartition modeling. For comparison samples, sources with size corresponding to a direct measurement of the largest linear size are shown as $\times$ or $+$, while sources with equipartition-inferred sizes are shown as filled circles or lines. For inferred equipartition sizes for AT2018hyz and AT2019azh, models assuming a spherical outflow are shown as solid lines, while those assuming a collimated jet are shown as dotted lines. The post-starburst sources have lower luminosities and smaller sizes than most of the AGN samples, though the post-starburst source luminosities overlap with the transient sample from \citet{Kunert2025}. The post-starburst sources have properties similar to the non-relativistic TDE samples, particularly those of \citet{Somalwar2025b, Goodwin2025}. 
}
\label{fig:size}
\end{figure*}

\begin{figure}
\begin{center}
\includegraphics[width=0.5\textwidth]{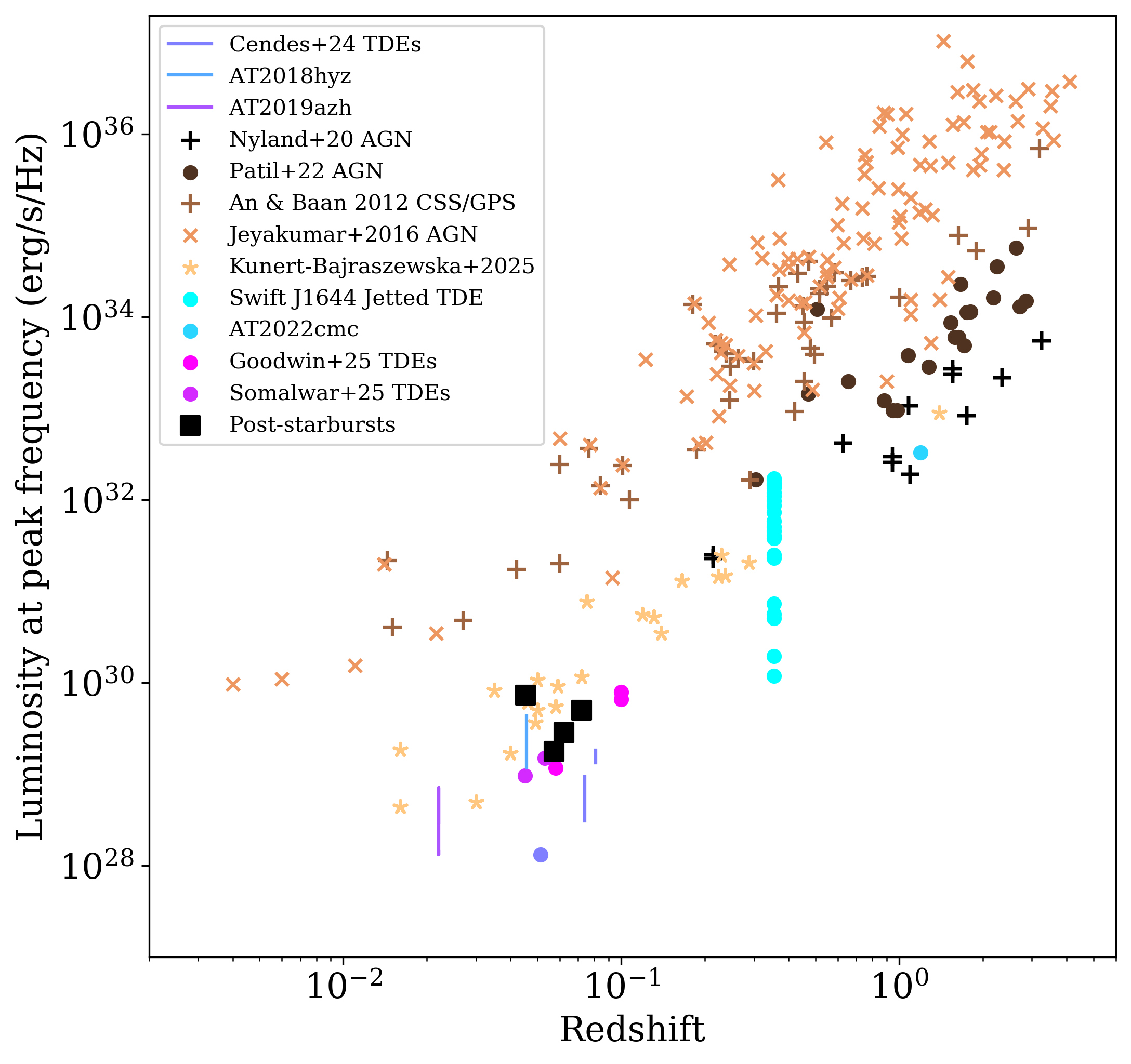}
\end{center}
\caption{Luminosity at peak frequency vs. redshift for the post-starburst sources and comparison samples, with plot points as in Figure \ref{fig:size}. While there is a clear trend with luminosity and redshift, the post-starburst sources (black squares) are among the lower luminosity sources, even at comparable redshifts.
}
\label{fig:Lz}
\end{figure}

\subsection{Comparison to AGN}
\label{sec:agn}

The radio sources we observe in this sample of post-starburst galaxies have similar SED shapes as the gigahertz peaked sources (GPS) and compact steep spectrum/compact symmetric sources (CSSs/CSOs) seen in some radio AGN \citep{Peacock1982, Fanti1990, Odea1998, Odea2021}. These classes of radio AGN have SEDs indicative of self-absorbed synchrotron emission, likely due to the compact nature of these sources. The radio variable sources found by \citet{Nyland2020} also show this spectral shape. 

We compare the peak luminosity, frequency, and size of peaked AGN and our post-starburst sources in Figure \ref{fig:size}.  We select peaked spectrum AGN from compilations by \citet{An2012}\footnote{\citet{Gregory1992, Patnaik1992, White1992, Condon1998, Stanghellini1998, T03, G05, T05, Gallimore2006, Healey2007, Orienti2007, Orienti2010, Orienti2012}} and \citet{Jeyakumar2016}\footnote{\citet{Fanti1990, Stanghellini1998, Fanti2001, Dallacasa2000, Peck2000, Snellen2002, Bolton2004, Edwards2004, Snellen2004, G05}}, as well as the obscured quasar sample from \citet{Patil2022}, and variable sources from \citet{Nyland2020} and \citet{Kunert2025}. While the peak frequencies of our sources are similar to the peaked spectrum AGN, the luminosities of the post-starburst radio sources are considerably lower. Our post-starburst sources are at lower redshift ($z\sim0.04-0.07$ vs. medians of $z\sim0.07-0.9$ for the peaked AGN samples; Figure \ref{fig:Lz}). Our lower-luminosity sources are more similar to the transient sources found by \citet{Kunert2025}, which were found by comparing NVSS \citep{Condon1998} and VLASS. 

The inferred sizes of our post-starburst sources are much smaller than the AGN source sizes, though we caution that we do not have the spatial resolution to measure the source sizes directly. The post-starburst sources are all unresolved in our VLA observations, but the Ku-band beam size of 0.12 arcsec can be used to place upper limits on the sizes of these sources to be $<100-160$ pc. The equipartition modeling described in \S\ref{sec:sed} infers smaller sizes of 0.03-0.09 pc. We compare these sizes to the peaked spectrum AGN from \citet{An2012} and \citet{Jeyakumar2016}, for which the linear size corresponds to a measured largest linear size. These sources scatter about the \citet{Odea1997} and \citet{Orienti2014} average scaling relation between source size and frequency. We also include the sample of AGN from \citet{Patil2022}, for which inferred equipartition sizes have been measured. These sizes are systematically lower than the largest linear size sources, likely due to the geometry of the sources. An extended source with axis ratios $3-10:1:1$ would appear as $2-5\times$ larger if the largest linear size were considered instead of the average radius from equipartition modeling. Thus, even given the uncertainties in the post-starburst size inference, they are significantly smaller than typical peaked spectrum AGN.

The small sizes and low luminosities of the post-starburst sources allow us to consider possible evolutionary paths relative to the peaked AGN. Our radio sources could be lower luminosity analogs of the peaked AGN, more similar to the faint end of the peaked sources seen by \citet{Kunert2025} \footnote{Although, both \citet{Kunert2025} and \citet{Readhead2024} suggest that many of the lower luminosity peaked sources may instead be TDEs, see continued discussion in \S\ref{sec:tdes}.}. If the post-starburst sources are lower luminosity analogs of typical peaked AGN, they are unlikely to evolve directly into the more luminous peaked sources on the evolutionary track of \citet{An2012}. Peaked sources evolve towards large size and lower peak frequency as they age, if the spectral turnover is driven by synchrotron self-absorption, a trend seen in both AGN \citep{An2012} and TDE \citep{Berger2012} sources. AGN will evolve along the size-peak frequency relation shown in Figure \ref{fig:size}. 
For the post-starburst sources to evolve into the peaked AGN sources, the post-starburst sources would have to grow in size without a corresponding decrease in the peak frequency, in contrast to the expected growth trends. Furthermore, if the post-starburst sources were evolving into the peaked AGN sources on the $<1000$ year timescales expected for the luminous AGN, we would expect to see a wider range of post-starburst radio luminosities. However, we do not see a large population of higher-luminosity radio sources in the post-starburst sample. The VLASS epoch 1 3 GHz luminosities of the four variable post-starburst sources are $\log L_\nu/(\rm{erg \ s^{-1} Hz^{-1}}) = 29.3-29.7$, typical of the median of the VLASS-detected post-starbursts ($\log L_\nu/(\rm{erg \ s^{-1} Hz^{-1}}) =29.6$; 50th percentile range $\log L_\nu/(\rm{erg \ s^{-1} Hz^{-1}}) =29.3-30.0$). 
Thus, if these post-starburst sources are driven by low luminosity AGN, the AGN must be on a parallel track of evolution, perhaps from a lower accretion rate. 
On much longer timescales of Gyr, post-starburst galaxies are expected to evolve into quiescent early type galaxies, and may eventually host luminous peaked source AGN.

\begin{figure*}
\begin{center}
\includegraphics[width=0.32\textwidth]{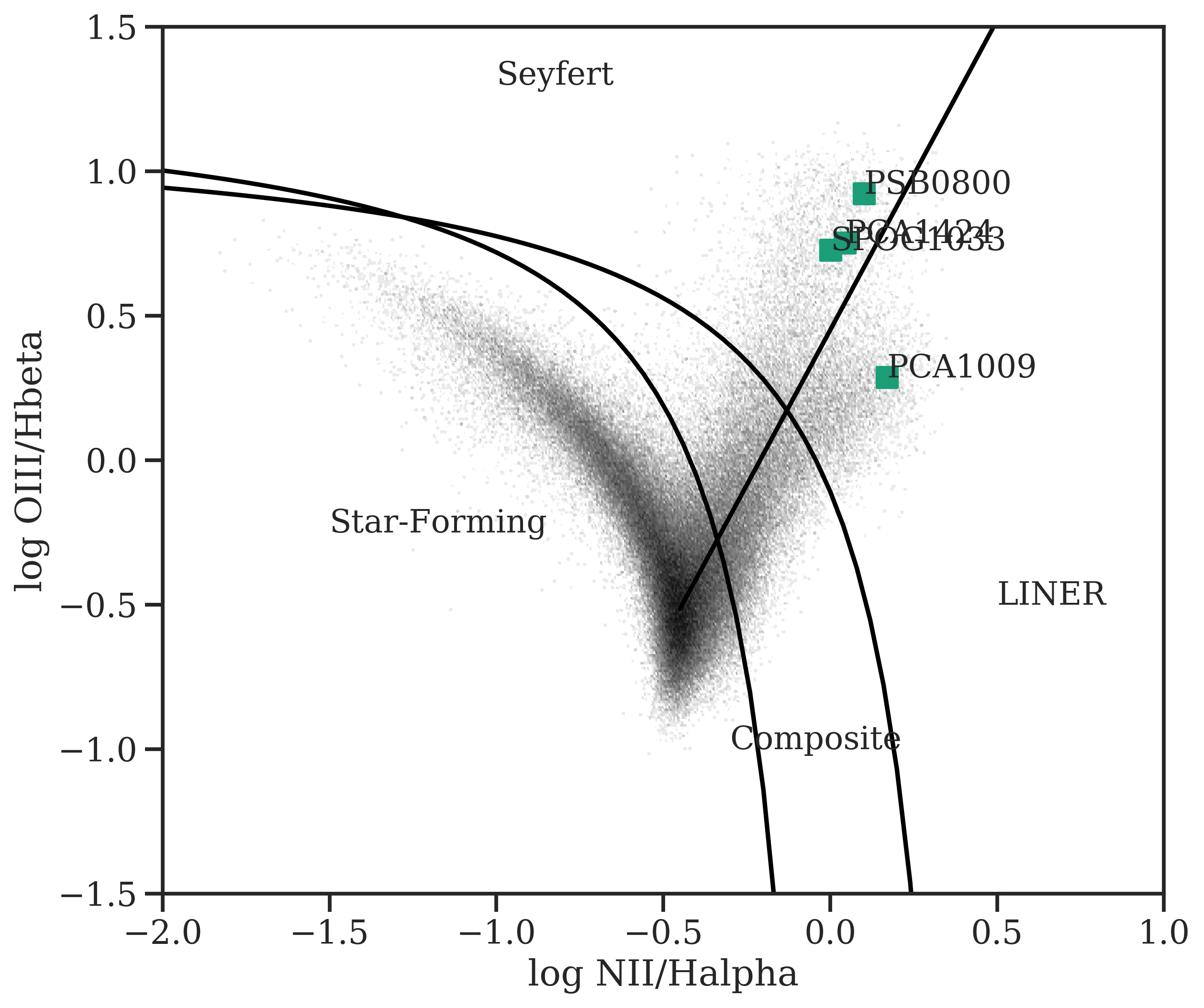}
\includegraphics[width=0.32\textwidth]{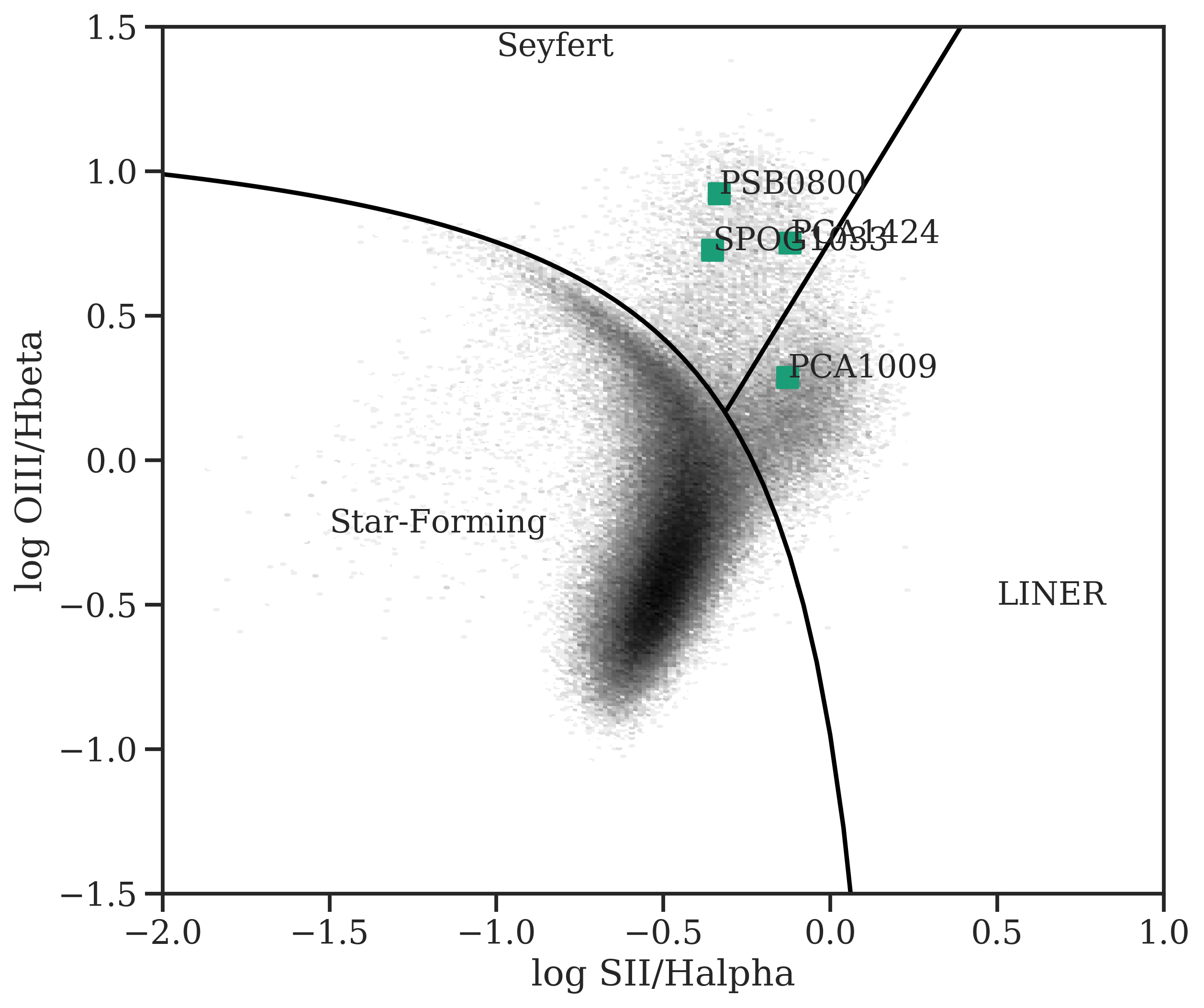}
\includegraphics[width=0.32\textwidth]{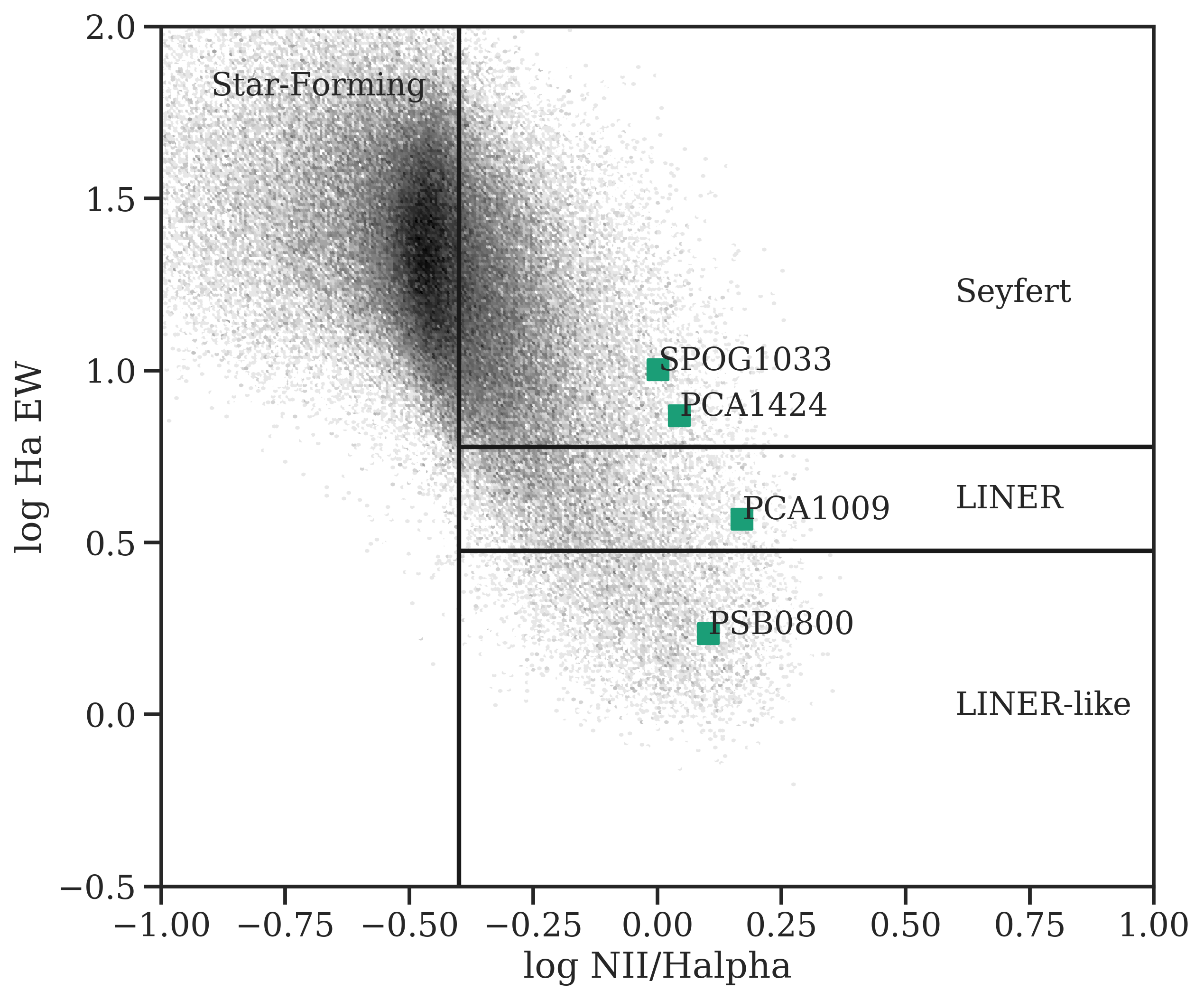}
\end{center}
\caption{BPT \citep{Baldwin1981,Kewley2001,Kauffmann2003} and WHAN  \citep{Cidfernandes2011} diagrams for our sources, using data from MPA-JHU galspec. Two sources (SPOG1033, PCA1424) are Seyferts, but the others are more LINER-like.
}
\label{fig:bpt}
\end{figure*}

Several of our post-starburst sources have other lines of evidence for weak AGN. In Figure \ref{fig:bpt} we show emission line ratio diagrams using the optical emission line data from the MPA-JHU catalog \citep{Kauffmann2003, Tremonti2004, Brinchmann2004}. Two sources (SPOG1033, PCA1424) are Seyferts across multiple classification methods, but the others are more LINER-like. One of the sources (PSB0800) also has an extended emission line region \citep{French2023b}. This could be from past AGN activity, or past TDE activity \citep{Wevers2024, Mummery2025}. SPOG1033 has a ROSAT detection that could be associated. The ROSAT source is 14\arcsec away from the optical center, consistent within the large $\sim25$\arcsec positional uncertainty from ROSAT. From the ROSAT count rate, the estimated X-ray flux is $1.4\times10^{-13}$ erg s$^{-1}$ cm$^{-2}$. If this source is at the distance of SPOG1033, the luminosity would be $L_{\rm X} = 1.3\times10^{42}$ erg s$^{-1}$. This luminosity is similar to the range of $L_{\rm X}$ for other SPOGs in the study by \citet{Lanz2022}. However, it is fainter than what we would expect given the \citet{Merloni2003} relation between $L_{\rm X}$, radio luminosity, and black hole mass. Assuming the black hole mass for this object is $10^7$ \Msun (consistent with the \citet{McConnell2013} relation for this object), this source should have $L_{\rm X}\sim 10^{44}$ erg~s$^{-1}$. 

If the radio sources in these post-starburst galaxies are driven by AGN, there are a number of possibilities for the observed radio variability. Extreme changes in the radio luminosity are expected for young, newly-launched jets \citep{An2012, Nyland2020}. Changes in doppler boosting from the realignment of jets, or interactions between jets and dense clumps of interstellar medium could also result in large-scale changes in luminosity on decades scales \citep{Nyland2020}. Other intermediate scenarios are possible, such as intermittent gas accretion. For example, clumpy gas accretion during the post-merger phase may fuel more intermittent AGN fueling. 

While the WISE lightcurve for PCA1424 is similar to some changing-look AGN as discussed in \S\ref{sec:lightcurves}, there are several lines of evidence against our sources being caused by changing-look AGN.  Some changing-look AGN with distinct flares, like 1ES 1927+654 \citep{Meyer2025} have been observed to have coindicent radio flares. However, \citet{Birmingham2025} find radio flares to be rare in a large sample of changing-look AGN. In a study of the host galaxy properties of changing-look AGN, \citet{Verrico2025} find that changing-look AGN are not often found in post-starburst galaxies, and that changing-look AGN have host galaxies similar to other types of AGN. Thus, in contrast with the TDEs we discuss in the following section, we do not expect to have a high rate of changing-look AGN in the post-starburst sample.

\subsection{Comparison to TDEs}
\label{sec:tdes}

Could the radio flares we see in the post-starburst galaxies be caused by TDEs instead of AGN activity? 
The rate of TDEs in post-starburst galaxies is enhanced by a factor of $\sim20-100\times$ relative to other galaxies \citep[e.g.,][]{Arcavi2014, French2016, French2020}. Scaling from a total rate of $3\times10^{-5}$ per galaxy per year \citep{Yao2023}, this implies a rate of $6\times10^{-4}-3\times10^{-3}$ per galaxy per year in post-starburst galaxies. A rough lower limit on the rate of the radio flares in the post-starburst sample is 4 per 5000 galaxies (our parent sample of galaxies) per 20 years (FIRST vs. VLASS), which is $4\times10^{-5}$ yr $^{-1}$. Even if only half of TDEs drive radio outflows, the TDE rate in these galaxies is high enough to explain the 1-4 radio flares we see, if the radio emission lasts $>1$ year. 

Jetted TDEs like Swift J1644 are rare, with expected rates only $\sim1$\% of the total TDE rate \citep{Alexander2020}, and it would be unlikely to have found these by chance in our sample. However, larger searches for radio variable nuclear sources may have found such events, see \citet{Chen2024}.

We compare the luminosities, peak frequencies, and inferred sizes of radio sources from this study to TDE radio flares in Figure \ref{fig:size}. We include the relativistic, jetted TDE Swift J1644 \citep{Berger2012, Zauderer2013, Eftekhari2018}, the jetted TDE AT2022cmc \citep{Pasham2023}, the late-time jetted TDE AT2018hyz \citep{Cendes2022}, the sample of optically-selected TDEs from \citet{Cendes2024}, the sample of radio-selected TDEs from \citet{Somalwar2025b}, the optically-selected TDE AT2019azh \citep{Goodwin2022}, and the sample of X-ray-selected TDEs from \citet{Goodwin2025}. We exclude sources with peak frequencies below the observed frequency range, for which turnover frequencies cannot be constrained. The post-starbursts have luminosities at the high end of the non-relativistic TDEs, with similar peak frequencies and sizes. 

The presence of an optical or IR flare is expected if the post-starburst sources are caused by TDEs. 
We do not see coincident optical flares in any of the four post-starburst targets (see \S\ref{sec:lightcurves}), but it is possible that we missed an optical flare due to a lack of optical coverage or sensitivity. 
Given the long delay times between the observed optical peaks of TDEs and the observed radio emission, the radio flares we see in $\sim2017-2023$ could have been caused by earlier TDEs with no optical coverage. Given the typical optical peak brightnesses of TDEs ($M_r \sim -18$-- $-22$; \citealt{Yao2023}), the peak flux for the distance of our targets should be $0.05-5$ mJy (14.5-19.5 mag). Fainter TDEs could have been missed, especially during the earlier surveys. Alternatively, an optical flare could have been missed due to dust obscuration (see discussion below), though we do not see strong evidence for heavy dust obscuration in these sources (see \S\ref{sec:origin}).

\begin{figure}
\begin{center}
\includegraphics[width=0.5\textwidth]{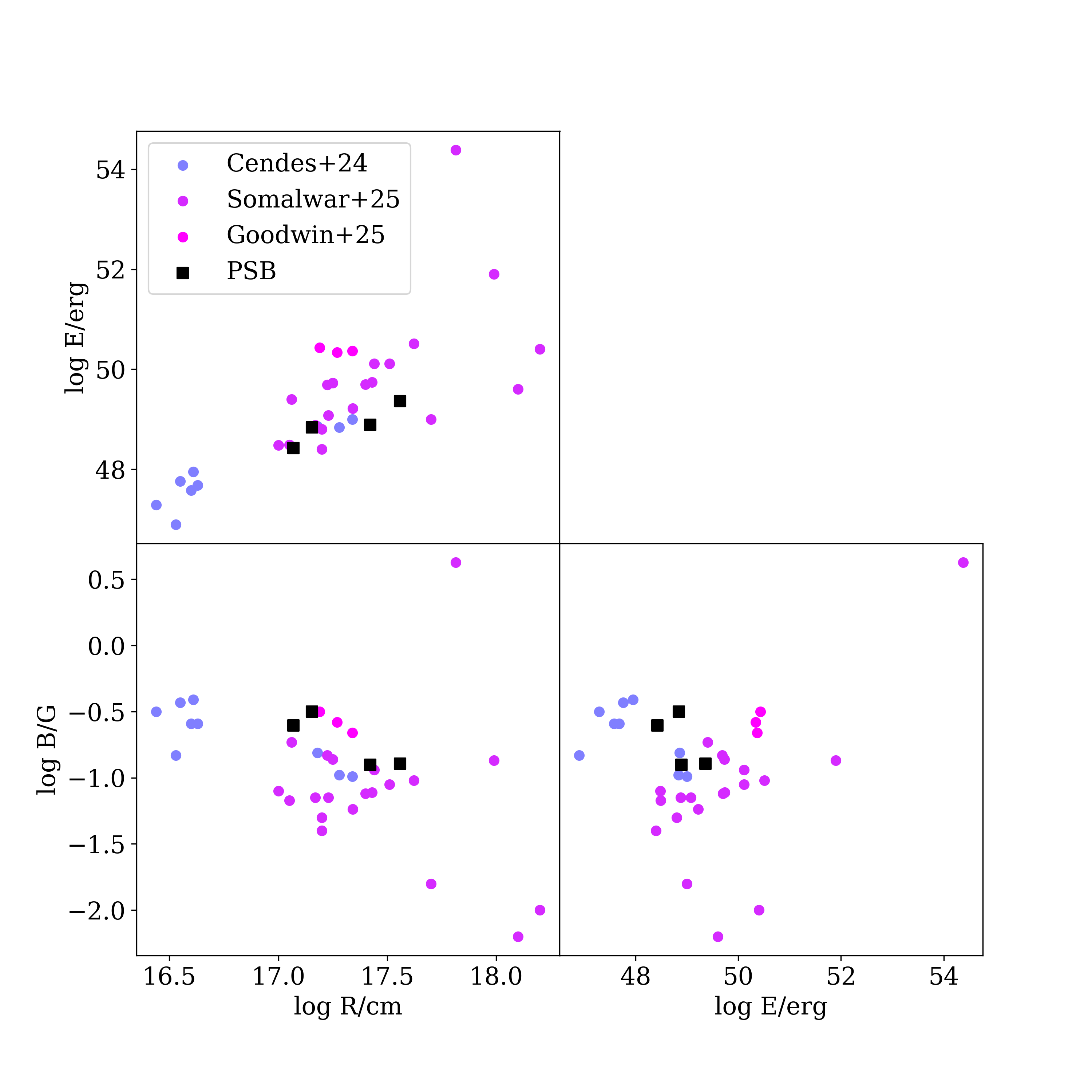}
\end{center}
\caption{Best-fit equipartition model parameters for the four post-starburst targets, as well as the TDE samples from \citet{Cendes2024}, \citet{Somalwar2025a}, and \citet{Goodwin2025}. The post-starburst sources have parameters broadly consistent with the scatter from all three TDE samples. 
} 
\label{fig:tde_equi}
\end{figure}

Differences in spectral shape can in principle be used to distinguish between TDEs and AGN. 
The SED shapes of TDEs tend to be fit by a single self-absorbed power law component, with the high frequency (optically thin) slope varying over a small range, and the low frequency (optically thick) component well fit by a slope of $\nu^{5/2}$. In contrast, the GPS/CSO sources have a wider variety of SED shapes, not consistent with a single self-absorbed synchrotron component \citep{delaParra2024}. Two of the post-starburst sources prefer a shallower low frequency slope than $\nu^{5/2}$, but our low frequency constraints are limited (as are those of many of the TDEs).  Lower frequency observations would be needed to search for deviations from emission from simple SSA models. We compare our inferred energies, sizes, and magnetic fields to the sample of TDEs fit in this way by \citet{Cendes2024, Somalwar2025b, Goodwin2025} in Figure \ref{fig:tde_equi}. Compared with the TDEs fit by \citet{Cendes2024}, our four sources are among the TDEs with higher radius and energy. Our post-starburst sources have typically higher energies than the \citet{Cendes2024} sample, but are broadly consistent with the scatter from all three TDE samples.

\begin{figure}
\begin{center}
\includegraphics[width=0.5\textwidth]{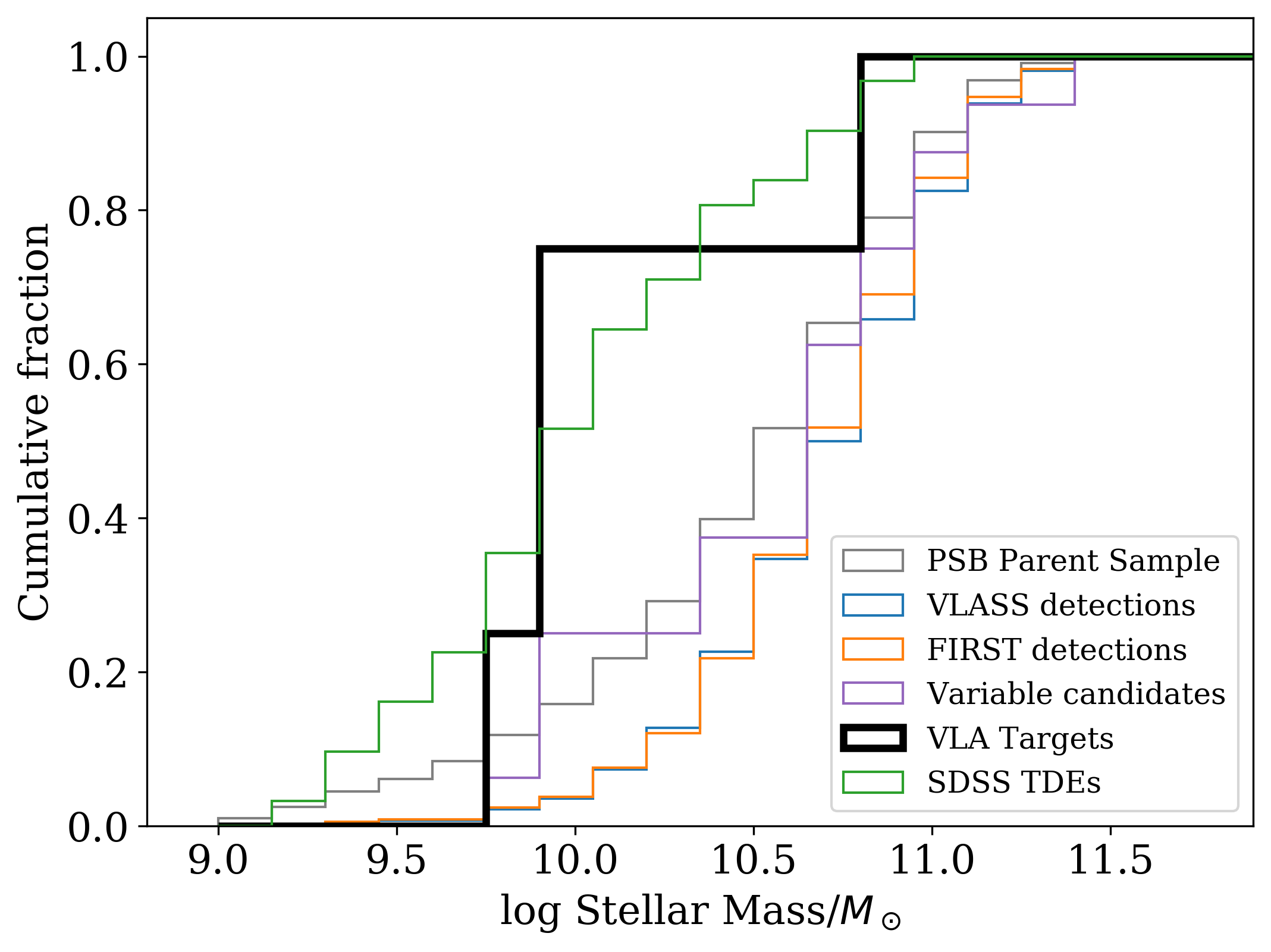}
\end{center}
\caption{Cumulative distribution of the stellar masses of (grey) the parent sample of 5000 post-starburst galaxies from the SDSS, (blue) the subset with VLASS detections, (orange) the subset with FIRST detections, (purple) the subset of 16 with VLASS detections and FIRST non-detections, which could potentially be variable sources, and (black) the subset of 4 definitively variable source we present here. The VLASS- and FIRST- detected post-starburst galaxies tend to have higher stellar masses than the parent sample (blue and orange vs. grey), consistent with an AGN source. We also compare to TDE host galaxies (green). The 4 variable sources we consider here have lower stellar masses, more consistent with the TDE host galaxies (black vs. green) than expectations from AGN. 
}
\label{fig:stmass}
\end{figure}

The distribution of stellar masses provides a population-scale test of whether the events we see are driven by TDEs or AGN.  We would expect radio AGN to be biased towards higher stellar mass \citep[e.g.,][]{Heckman2014}, and TDEs to have intermediate stellar masses, due to the fall-off at high stellar mass from event horizon capture (at $M_{BH}>10^8$ \Msun) and the peak TDE rate at black hole masses $\sim10^6 \ M_\odot$ \citep{Yao2023,Chang2025}. In Figure \ref{fig:stmass}, we show the cumulative fractions of stellar mass for our parent post-starburst galaxy sample, as well as radio-detected subsamples. We also compare to the distribution of stellar masses from 34 TDE host galaxies with SDSS spectra (updated from \citet{French2020}). The subset of post-starburst galaxies with either VLASS or FIRST detections tend to have higher stellar masses than the parent sample. This is consistent with most of these detections being from AGN. In contrast, the four VLA targets have a lower stellar mass distribution than the parent sample, more consistent with those typical of TDE host galaxies, though we caution that this is a small sample.

Thus far, the enhanced TDE rate has been seen for E+A galaxies (like our target PSB0800; selected with cuts against H$\alpha$ emission to select against current star formation), as well as quiescent Balmer-strong galaxies (selected using similar cuts against H$\alpha$ emission, yet with less strict requirements for Balmer absorption), but not in post-starburst samples like the SPOG and PCA samples we consider here, which tend to have higher residual star formation rates. One reason we may not see the same over-representation in the SPOG and PCA samples is if these galaxies have more dust obscuration, which does not obscure radio emission. IR and radio flares have been used to search for dust-obscured TDEs. Several TDEs have been identified to have IR or radio only flares \citet{Mattila2018,Jiang2021, Reynolds2022,Masterson2024,Dykaar2024}. While some IR flares have been found in starburst galaxies \citep{Mattila2018, Reynolds2022}, dedicated searches for IR flares have found their hosts to be similar to AGN hosts \citep{Dodd2023} or distributed with normal, non-starbursting galaxies \citep{Masterson2024}. AGN variability is difficult to disentangle from individual TDEs without real time optical spectroscopy, though mid-IR spectroscopy may be able to shed light on the presence and rate of dust-obscured TDEs \citep{Masterson2025}.

If the radio outflows we observe are driven by TDEs, there are also a number of intermediate possibilities aside from the total disruption of a star. Outflows could also be launched by partial TDEs \citep{Makrygianni2025}, the return of unbound material in dense circumnuclear environments \citep{Ryu2024}, or mass transfer from tidally-heated stars \citep{Yao2025}. 

We note here that the TDEs AT2018hyz and AT2019azh are both in E+A post-starburst hosts like PSB0800, although they did not meet our selection cuts on FIRST, VLASS epoch 1 change. 
Radio flares that are discovered in real time and followed-up with multiwavelength observations, such as those in AT2018hyz and AT2019azh, will help to characterize the range of low-luminosity accretion events in post-starburst galaxies, to determine whether the primary driver is AGN or TDEs.

\subsection{Impact of Energy on the Host Galaxies}

Regardless of whether the origin of the accretion is a tidally-disrupted star or gas inflows, we can use the observed rate of these radio flares and their energies to constrain the impact on the host galaxy. While AGN feedback traditionally assumes a persistent accretion disk, \citet{Smercina2022} and \citet{Wevers2024a} have explored the possibility of TDE-driven feedback.

The total binding energy of the molecular gas from resolved CO observations \citep{Smercina2022} is of order $E \sim (10^8 \ M_\odot)(100 \ \rm km \ s^{-1})^2 \sim 2\times10^{55}$ erg. \citet{Smercina2022} estimate that the energy required to replenish the entire turbulent energy is $\sim10^{55}$ erg. The kinetic power in the observed low velocity molecular outflows seen by \citet{French2023a} is $\sim 10^{41}$ erg/s. Most of the mass in outflows is expected to be in the molecular phase, and the ionized gas outflows seen in post-starburst galaxies contain significantly less energy \citep{Fodor2025}. \citet{Baron2022} find outflow kinetic power of $\sim10^{41}$ erg/s in ionized gas and $\sim10^{42}$ erg/s in neutral atomic gas in a sample of post-starburst galaxies, though this sample was selected without strict cuts against ongoing star formation and may represent an earlier phase of obscured but declining star formation. 

We consider two methods to estimate the energy available using the radio observations. The first method is to consider the minimum equipartition energies fit in \S\ref{sec:sed}, which are typically $\sim10^{49}$ erg. The total energy injected depends on the rate and lifetime of these flares, but this energy is so low that even assuming a rate $100\times$ higher than what we observe, it would take nearly a Gyr of sustained power, incompatible with the low radio duty cycle we observe. 

Alternatively, we can use the expected scaling between kinetic power and radio luminosity, following \citet{Zakamska2014, Smith2020} to be $L_{\rm wind}/(\nu L_{\nu, 1.4 \ GHz}) \sim 3.6\times10^{-5}$. At our measured luminosities, this infers kinetic power of $10^{42}-10^{43}$ erg/s from outflows in these galaxies. If this is sustained for $10^5$ years, comparable to the AGN duty cycle in other sources \citet{Lintott2009, Schawinski2015, Keel2017, Shen2021}, it would result in $3\times10^{54}-3\times10^{55}$ erg injected into the interstellar medium. These outflow energies are more than sufficient to power the molecular gas outflows that have been observed in post-starburst galaxies, and are capable of driving the observed turbulence and potentially unbinding the gas. Several studies of radio quiet AGN have been found to show clear impacts of radio jet activity on the molecular gas in the host galaxy \citep{Murthy2019, Venturi2021, Girdhar2022, Murthy2022}, although even these low-luminosity cases are typically more luminous than our post-starburst sources.

Thus, the predicted wind power of $10^{42}-10^{43}$ erg/s is enough to generate the required energy in only $10^5$ years. The duty cycle estimates from extended emission line region measurements by \citet{French2023b} of $\sim 5$\% allow for $\sim15$ Myr of time when the AGN is ``on" over the course of $\sim300$ Myr during the post-starburst phase, which provides ample time even if the actual wind power is relatively low. We note that the radio luminosities of these sources are $\sim10-100\times$ lower than the sources considered by \citet{Zakamska2014} and \citet{Smith2020}. It remains uncertain why the low luminosity sources we consider here are capable of maintaining quiescence, while higher power sources are often found in star forming galaxies.

\section{Conclusions}
\label{sec:conclusions}

We observed four post-starburst galaxies with variable radio emission, by selecting sources that were not detected in FIRST, yet detected in VLASS epoch 1. We obtained quasi-simultaneous followup with VLA in 5 bands (L, S, C, X, Ku; 1-18 GHz) to study the launching of outflows or jets during the period of galaxy evolution where galaxies have recently quenched. Our main conclusions are as follows:

\begin{itemize}
\item All four sources have risen in 1-2 GHz since FIRST, by factors of 2.5-25. Two of the four sources have clearly risen since the first epoch of VLASS, while the other two have roughly constant 3 GHz fluxes (Figure \ref{fig:lc_radio}). 

\item All four sources show peaked SEDs (Figure \ref{fig:equi}). These spectra are consistent with self-absorbed synchrotron emission. We fit each spectrum with an equipartition model to estimate the minimum energy, radius, and magnetic field.  

\item Two galaxies show IR flares. PSB1009 shows a clear flare, which was also picked up by other searches \citep{Assef2018,Wang2018}. PCA1424 also shows a rise over time. The WISE color is redder during the brighter epochs in both cases. There are no significant optical flares, variability, or changes.

\item These sources show similarities with both peaked radio AGN and the outflows launched from TDEs. The luminosities and sizes of our sources are smaller than most peaked radio AGN, but similar to the late-time radio TDEs. The expected TDE rate, particularly since it will be enhanced in post-starburst galaxies, is sufficient to explain some or all of the radio flares seen here. While we cannot definitively say whether these flares are driven by persistent AGN activity or individual TDEs, the stellar mass distribution of the hosts of our sources is more consistent with TDEs than AGN.

\item Regardless of whether the origin of the accretion energy is AGN or TDEs, the energy expected from these outflows is sufficient to drive the low velocity molecular gas outflows seen in some post-starburst galaxies. If this energy can be sustained for $10^5$ years, it would be enough to provide the observed turbulence seen in post-starburst galaxy molecular gas observations and unbind the observed gas reservoirs. 
    
\end{itemize}

\vspace{0.5cm}

K.D.F. acknowledges support from NSF grant AST 23-07375. 

Basic research in radio astronomy at the U.S. Naval Research Laboratory is supported by 6.1 Base Funding.

N.E., M.S., M.E.V. and K.D.F. acknowledge support from NSF grant AST 22-06164. M.S. and M.E.V. acknowledge support from the Illinois Space Grant Consortium. N.E. and M.E.V. acknowledge support from the Center for Astrophysical Surveys Graduate Fellowship.

K.R. acknowledges support from the NASA Astrophysics Data Analysis Program (ADAP) under grant 80NSSC23K0495.

The National Radio Astronomy Observatory is a facility of the National Science Foundation operated under cooperative agreement by Associated Universities, Inc.

This research has made use of the CIRADA cutout service at URL cutouts.cirada.ca, operated by the Canadian Initiative for Radio Astronomy Data Analysis (CIRADA). CIRADA is funded by a grant from the Canada Foundation for Innovation 2017 Innovation Fund (Project 35999), as well as by the Provinces of Ontario, British Columbia, Alberta, Manitoba and Quebec, in collaboration with the National Research Council of Canada, the US National Radio Astronomy Observatory and Australia’s Commonwealth Scientific and Industrial Research Organisation.

This publication makes use of data products from the Wide-field Infrared Survey Explorer, which is a joint project of the University of California, Los Angeles, and the Jet Propulsion Laboratory/California Institute of Technology, funded by the National Aeronautics and Space Administration.

Data from ZTF is based on observations obtained with the Samuel Oschin Telescope 48-inch and the 60-inch Telescope at the Palomar
Observatory as part of the Zwicky Transient Facility project. ZTF is supported by the National Science Foundation under Grants
No. AST-1440341 and AST-2034437 and a collaboration including current partners Caltech, IPAC, the Oskar Klein Center at
Stockholm University, the University of Maryland, University of California, Berkeley , the University of Wisconsin at Milwaukee,
University of Warwick, Ruhr University, Cornell University, Northwestern University and Drexel University. Operations are
conducted by COO, IPAC, and UW.

The Pan-STARRS1 Surveys (PS1) and the PS1 public science archive have been made possible through contributions by the Institute for Astronomy, the University of Hawaii, the Pan-STARRS Project Office, the Max-Planck Society and its participating institutes, the Max Planck Institute for Astronomy, Heidelberg and the Max Planck Institute for Extraterrestrial Physics, Garching, The Johns Hopkins University, Durham University, the University of Edinburgh, the Queen's University Belfast, the Harvard-Smithsonian Center for Astrophysics, the Las Cumbres Observatory Global Telescope Network Incorporated, the National Central University of Taiwan, the Space Telescope Science Institute, the National Aeronautics and Space Administration under Grant No. NNX08AR22G issued through the Planetary Science Division of the NASA Science Mission Directorate, the National Science Foundation Grant No. AST-1238877, the University of Maryland, Eotvos Lorand University (ELTE), the Los Alamos National Laboratory, and the Gordon and Betty Moore Foundation.

This work has made use of data from the Asteroid Terrestrial-impact Last Alert System (ATLAS) project. The Asteroid Terrestrial-impact Last Alert System (ATLAS) project is primarily funded to search for near earth asteroids through NASA grants NN12AR55G, 80NSSC18K0284, and 80NSSC18K1575; byproducts of the NEO search include images and catalogs from the survey area. This work was partially funded by Kepler/K2 grant J1944/80NSSC19K0112 and HST GO-15889, and STFC grants ST/T000198/1 and ST/S006109/1. The ATLAS science products have been made possible through the contributions of the University of Hawaii Institute for Astronomy, the Queen’s University Belfast, the Space Telescope Science Institute, the South African Astronomical Observatory, and The Millennium Institute of Astrophysics (MAS), Chile.

\software{Astropy \citep{astropy2013, astropy2018}, Matplotlib \citep{matplotlib}, NumPy \citep{numpy}}

\facility{VLA}


\bibliographystyle{aasjournal}
\bibliography{references}

\appendix{}
\section{Optical lightcurves}
\label{sec:appendix}

\begin{figure*}[h]
\begin{center}
\includegraphics[width=0.49\textwidth]{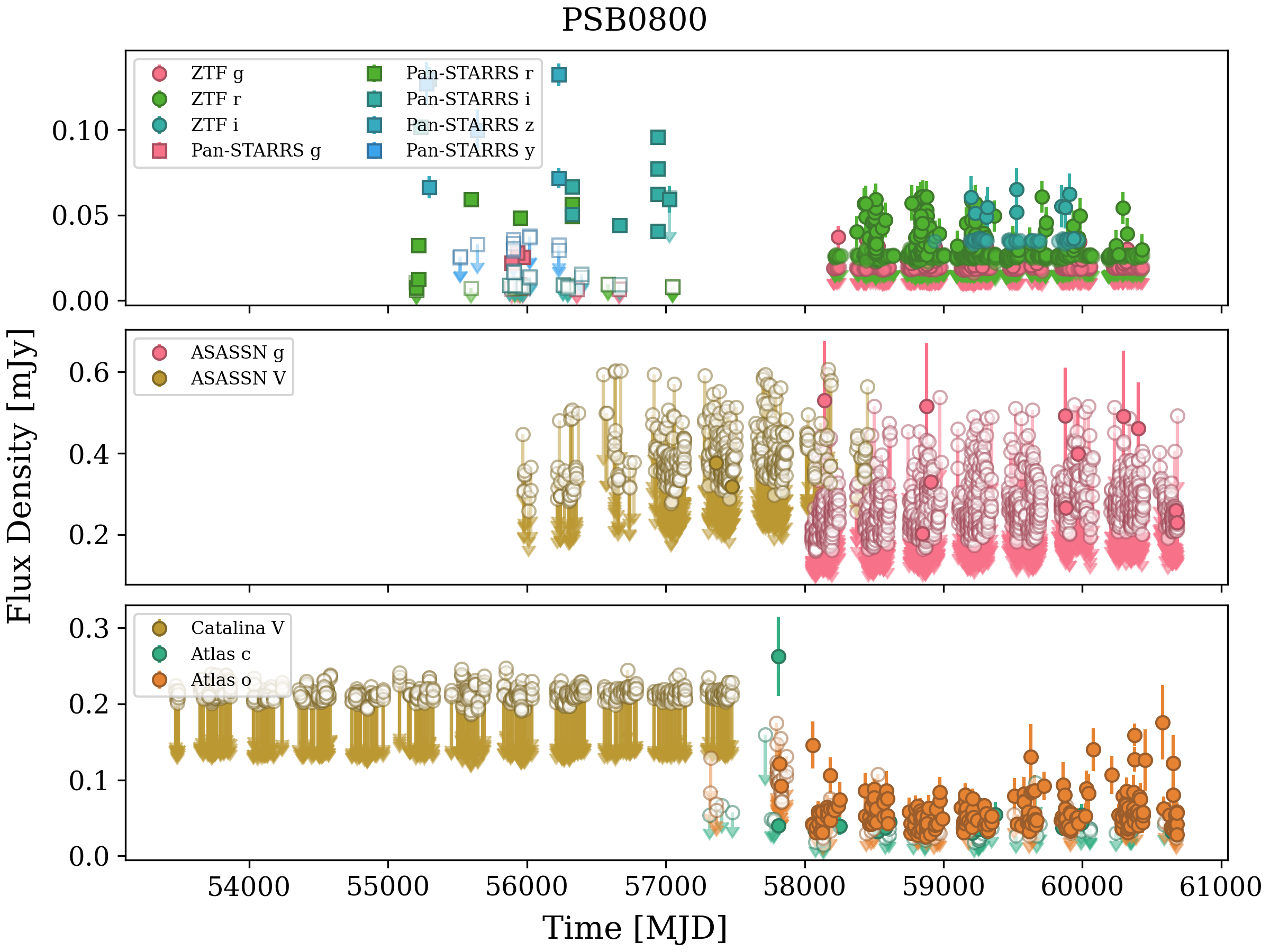}
\includegraphics[width=0.49\textwidth]{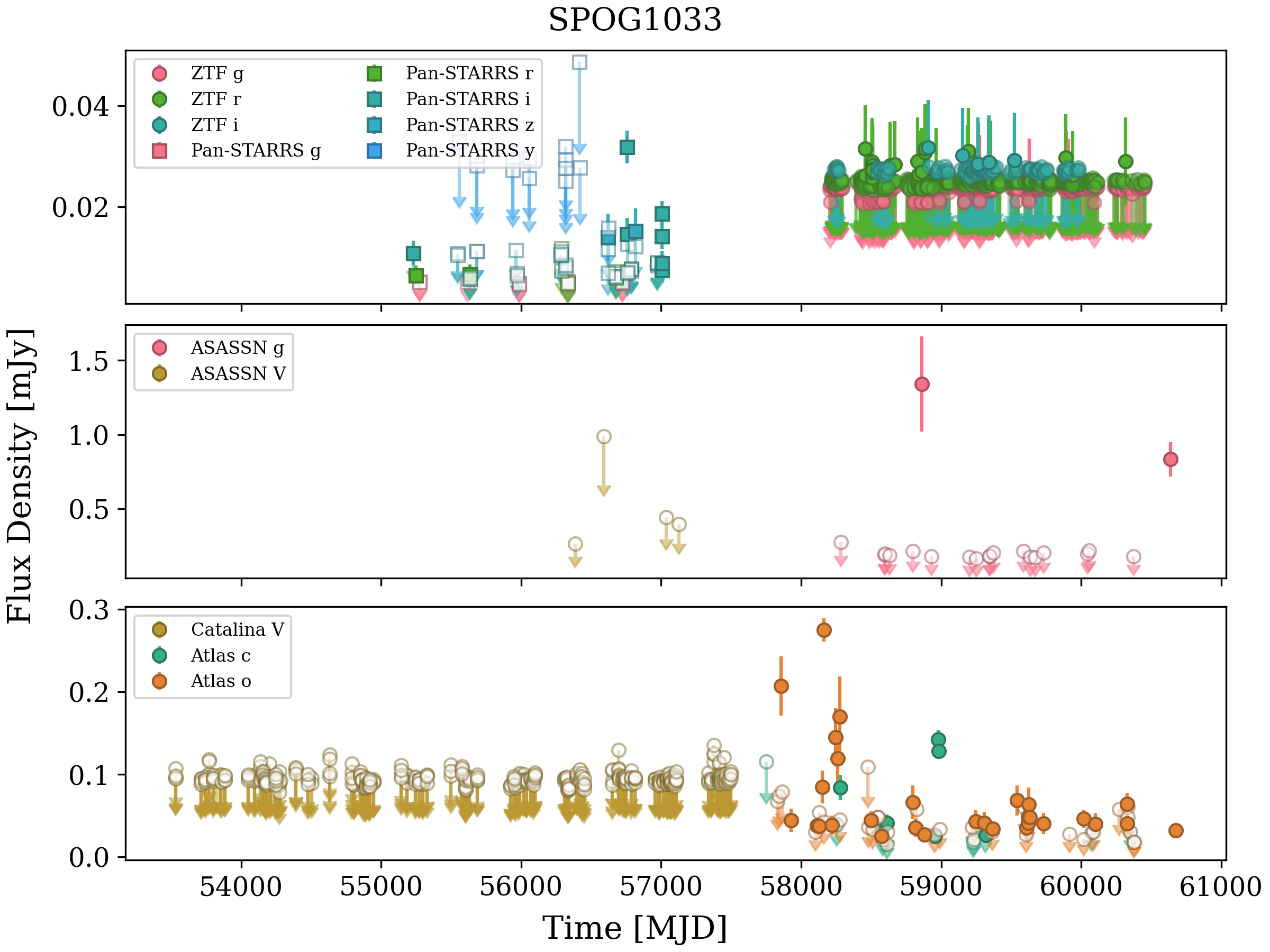}
\includegraphics[width=0.49\textwidth]{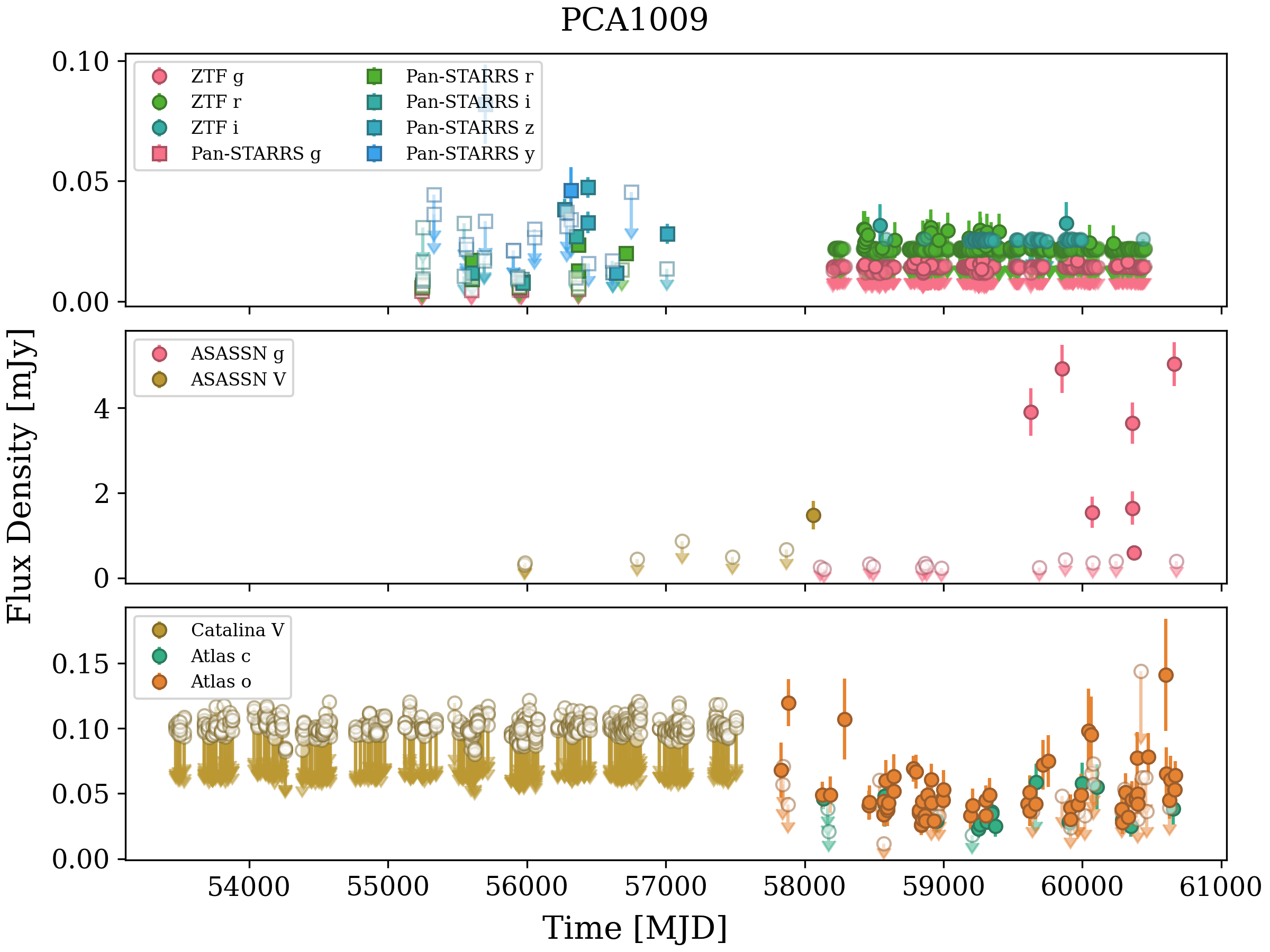}
\includegraphics[width=0.49\textwidth]{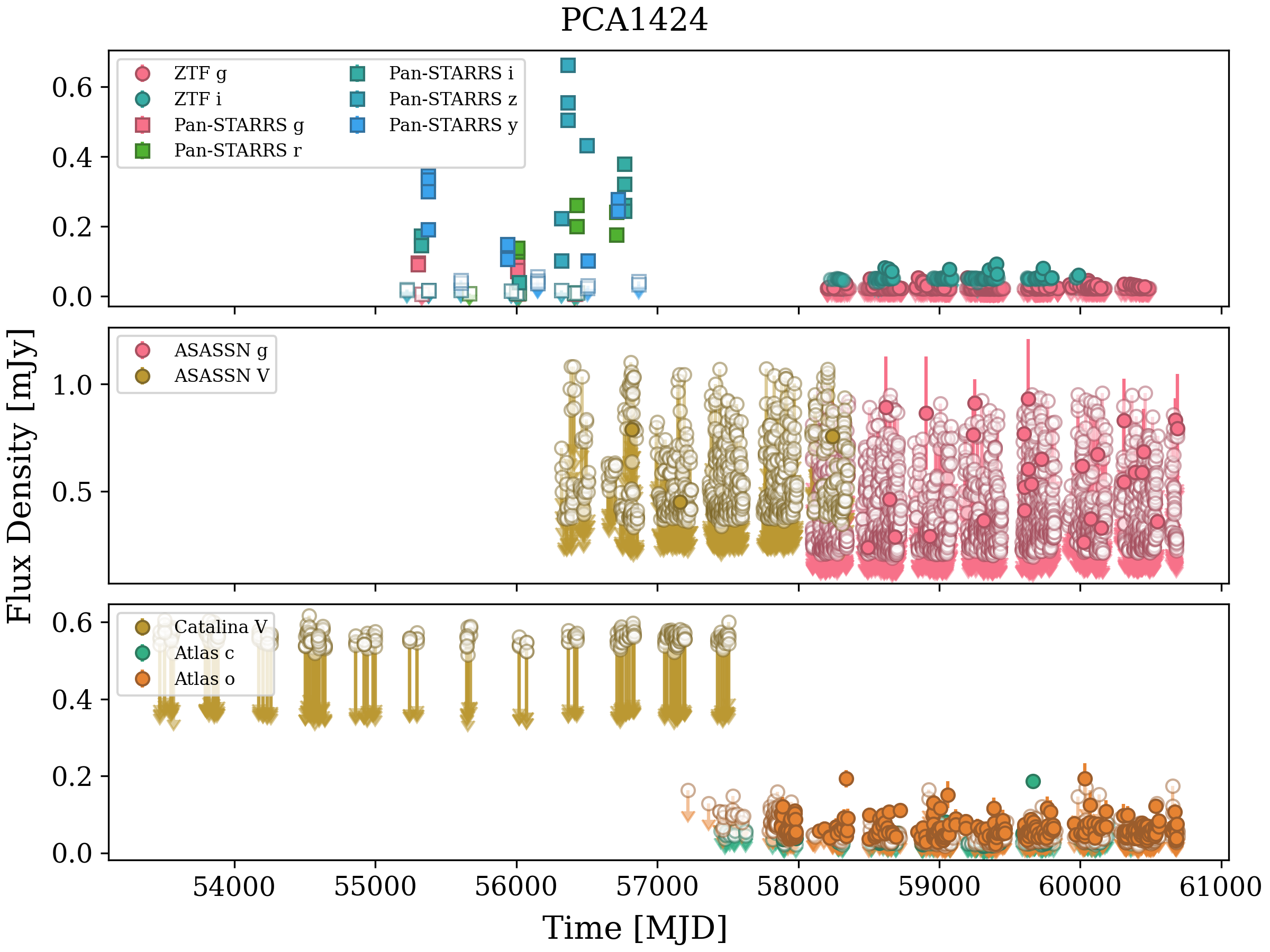}
\end{center}
\caption{Optical lightcurves for each source, as described in \S\ref{sec:lightcurves}. All upper limits shown are at the 3$\sigma$ level. No clear flares or variability are detected in the optical for any of these datasets. However, we note that it is possible that an optical transient was missed prior to the first observations we consider here or was dust-obscured.
}
\label{fig:optical}
\end{figure*}

\end{document}